\documentclass[11pt]{article}
  
\usepackage{amsmath,amssymb}

\usepackage{epsfig}  
\usepackage{graphicx}               % Standard graphics package  
\usepackage{url}
\usepackage{hyperref}

\newcommand{\nc}{\newcommand}  

\nc{\beq}{\begin{equation}}  
\nc{\eeq}{\end{equation}}  
\nc{\beqa}{\begin{eqnarray}}  
\nc{\eeqa}{\end{eqnarray}}  
\nc{\bea}{\begin{eqnarray}}  
\nc{\eea}{\end{eqnarray}}  
\nc{\ra}{\rightarrow}  
\nc{\lsim}{\begin{array}{c}\,\sim\vspace{-21pt}\\< \end{array}}  
\nc{\gsim}{\begin{array}{c}\sim\vspace{-21pt}\\> \end{array}}  
\nc{\Tr}{{\rm Tr}}
\nc{\slsh}{\slash\hspace*{-0.22cm}}

\def\be{\begin{equation}}
\def\ee{\end{equation}}
\def\bea{\begin{eqnarray}}
\def\eea{\end{eqnarray}}
\def\bit{\begin{itemize}}
\def\eit{\end{itemize}}

\newcommand{\fourzero}{{4\hspace{-0.8mm}\oplus\hspace{-0.8mm}0}}
\newcommand{\threeone}{{3\hspace{-0.8mm}\oplus\hspace{-0.8mm}1}}
\newcommand{\twotwo}{{2\hspace{-0.8mm}\oplus\hspace{-0.8mm}2}}
\newcommand{\threezeroISR}{{3\hspace{-0.8mm}\oplus\hspace{-0.8mm}0\hspace{-0.8mm}\oplus\hspace{-0.8mm}\mbox{ISR}}}
\newcommand{\twooneISR}{{2\hspace{-0.8mm}\oplus\hspace{-0.8mm}1\hspace{-0.8mm}\oplus\hspace{-0.8mm}\mbox{ISR}}}
\newcommand{\missET}{\slash{\hspace{-2.5mm}E}_T} 

\title{  
\vspace*{-2.3cm}  
\begin{flushright}  
\normalsize{  
SLAC-PUB-14328 
  }  
\end{flushright}  
\vspace{1.5cm}  
\Large  
\textbf{
Identifying Dark Matter Event Topologies at the LHC
 \\
}\vspace*{1.0cm}   
}

\author{Yang Bai$^a$
and Hsin-Chia Cheng$^{b}$ 
\vspace{5mm}
\\
${}^{a}$\normalsize\emph{Theoretical Physics Department, SLAC, Menlo Park, CA 94025, USA} \vspace{1mm} \\
${}^{b}$\normalsize\emph{Department of Physics, University of California, Davis, CA 95616, USA}
}

\date{}

\begin{document}  
\setcounter{page}{0}  
\maketitle  

\vspace*{1cm}  
\begin{abstract} 
Assuming dark matter particles can be pair-produced at the LHC from cascade decays of heavy particles, we investigate strategies to identify the event topologies based on the kinematic information of final state visible particles. This should be the first step towards measuring the masses and spins of the new particles in the decay chains including the dark matter particle. As a concrete example, we study in detail the final states with 4 jets plus missing energy. This is a particularly challenging scenario because of large experimental smearing effects and no fundamental distinction among the 4 jets. Based on the fact that the invariant mass of particles on the same decay chain has an end point in its distribution, we define several functions which can distinguish different topologies depending on whether they exhibit the end-point structure. We show that all possible topologies ({\it e.g.,} two jets on each decay chain or three jets on one chain and the other jet on the other chain, and so on) in principle can be identified from the distributions of these functions of the visible particle momenta.  We also consider cases with one jet from the initial state radiation as well as off-shell decays. Our studies show that the event topology may be identified with as few as several hundred signal events after basic cuts. The method can be readily generalized to other event topologies. In particular, event topologies including leptons will be easier because the end points are expected to be sharper and there are more distinct invariant mass distributions from different charges.
\end{abstract}  
\thispagestyle{empty}  
\newpage  
  
\setcounter{page}{1}

\baselineskip18pt   

%%%%%%%%%%%%%%%%%%%%%%%%%%%%%%%%
\section{Introduction}
\label{sec:intro}
The Large Hadron Collider (LHC) have started taking data and many beyond the standard model (SM) theories will be tested. Many of these new models contain dark matter candidates whose stability is protected by some unbroken $\mathbb{Z}_2$ symmetry. Examples are the $R$-parity in the supersymmetric model~\cite{Dimopoulos:1981zb} and the KK-parity in models with Universal Extra Dimensions (UEDs)~\cite{Appelquist:2000nn,Cheng:2002iz,Cheng:2002ab}. In these models, the dark matter particle is the lightest $\mathbb{Z}_2$-odd particle and may be produced in pairs at the LHC via cascade decays of heavier $\mathbb{Z}_2$-odd particles.  At colliders, the dark matter particles escape the detectors and result in missing energies in the events. Knowing the missing particle mass and its interactions with the SM particles is crucial for testing whether it can indeed be the major component of the dark matter in the universe.

Because only the transverse part of the sum of the missing particle momenta can be measured at hadron colliders, the kinematics can not be fully reconstructed on an event-by-event basis. It is a challenging task to measure the dark matter particle mass at hadron colliders. Nonetheless, there have been a lot of progress in developing new techniques to determine the dark matter particle mass at hadron colliders despite the partially unknown dark matter particle momenta (see \cite{Barr:2010zj} for a review). On a single decay chain, one can use the end points  of visible particle invariant mass distributions to obtain relations among the masses of the particles on the same decay chain~\cite{Paige:1996nx,Hinchliffe:1996iu,Allanach:2000kt,Gjelsten:2004ki,Gjelsten:2005aw,Miller:2005zp,Nojiri:2008vq,Burns:2009zi,Matchev:2009iw}.  If one uses both decay chains, the end point of the variable $M_{T2}$ can also be used to determine missing particle masses~\cite{Lester:1999tx,Barr:2003rg,Lester:2007fq,Cho:2007qv,Barr:2007hy,Cho:2007dh,Cheng:2008hk,Nojiri:2008hy,Burns:2008va,Bai:2009it,Alwall:2009zu,Konar:2009wn,Konar:2009qr}. If the decay chains are long enough and the event topology is known, one can even solve the kinematics of the full system by combining events~\cite{Kawagoe:2004rz,Cheng:2007xv,Cheng:2008mg,Cheng:2009fw,Nojiri:2010dk}. 

In almost all of those analyses, the event topology has been assumed to be known from the beginning. However, it frequently happens that the same final states can come from different event topologies arising from the same model or different models. If a wrong topology is assumed, these mass determination methods will not give sensible results. Therefore, the \emph{first} step towards measuring the dark matter mass should be identifying the dark matter event topology. In this paper, we are trying to clear the topology ambiguities and close the gap between the real experimental data and the different dark matter mass measurement techniques. 

For a given final state, there are discrete choices of corresponding event topologies. The kinematic distributions are in general functions of the masses, spins and couplings of all particles in the process, in addition to the event topology. Therefore, to distinguish topologies we would like to use binary information in these distributions which is independent of the details of the model. An important observation is that the kinematics of the visible particles from a single decay chain is constrained by the mother particle mass, and hence the invariant mass distributions of these visible particles will have end points. On the other hand, the invariant mass combinations of particles from different chains are not constrained (except by the center-of-mass energy) and are not expected to show the end-point structure. Using these facts we can design functions of the invariant mass combinations of the visible particles to exhibit end points for certain topologies but not for the others. The correct topology can emerge if there are enough such functions to differentiate all possible topologies.

If there are different types of visible particles in the event sample (for example, there are both jets and leptons or leptons with different flavors and different charges), many different kinds of invariant mass combinations can be formed. By examining whether each of them has an end point or not, one can easily tell if these objects in the corresponding invariant mass combination belong to the same decay chain. The task is more difficult if all the visible objects are jets. Not only there are fewer different types of invariant mass combinations, but the jets also suffer more from the experimental smearing effects. In this paper, rather than exhausting all possible final states we choose the quite challenging 4 jets plus missing transverse energy (MET) scenario as a case study.  This final state can have a large production cross section at the LHC and a good discovery chance even with early LHC data~\cite{Izaguirre:2010nj}~\cite{Alves:2010za}. Without considering the case in which two jets are from $W$ or $Z$ boson decays, there are three different topologies for $4 j + \missET$ even without the contamination of the initial state radiation (ISR): two jets on each decay chain ($\twotwo$); three jets on one chain and the other jet on the second chain ($\threeone$); all four jets on a single chain and only the dark matter particle on the other chain ($\fourzero$). The $\fourzero$ topology in principle can be identified by looking at the invariant mass distribution of all four jets, since only $\fourzero$ can have an end point under this function. To identify the other cases one needs to cleverly define certain functions of the 2-jet or 3-jet invariant masses which can preserve the end-point structure given that we do not know which 2 jets or 3 jets to use {\it a priori}.

In our detailed study, we consider the case that
the four jets have similar energies so that they can not be divided into different groups based on their $E_T$'s. We include the detector effects on the simulations as well as the initial and final state radiations to make the study more realistic. Furthermore, we will consider two other cases which give rise to the same final state: two jets on one chain and one jet on the other chain with the fourth jet coming from ISR; three jets on one chain and zero jet on the other chain with the fourth jet from ISR. We will show that with the help of two additional functions those two topologies may also be distinguished. 

Our paper is organized as follows. In Section~\ref{sec:fourjet}, we first define four invariant-mass functions for distinguishing the $4 j + \missET$ event topologies based on some theoretical motivations, and we perform the detailed parton-level studies to show that they indeed exhibit the behaviors that are expected.   In Section~\ref{sec:particle} we perform the realistic particle-level analysis on these functions. Even though the showering, hadronization, and the detector smearing make the end points harder to identify, we show that by fitting the slopes of the distributions and comparing the orders we can still distinguish  different topologies. We also study the two ISR cases: $\twooneISR$ and $\threezeroISR$ in Section~\ref{sec:ISR} and off-shell decays in Section~\ref{sec:Off-shell}. The cases with mixtures of topologies are considered in Section~\ref{sec:mix}. We then discuss strategies to identify event topologies for other final states and conclude our paper in Section~\ref{sec:Discussions}.

%%%%%%%%%%%%%%%%%%%%%%%%%%%%%%%%
\section{Kinematic Functions for $4j + \missET$ Event Topologies Identification}
\label{sec:fourjet}

In this section, we construct functions of visible particle momenta which can be useful to identify event topologies. We use the $4j+\missET$ final state for a detailed case study. Similar functions can easily be constructed for other final states. As motivated by the SUSY-like models, we consider that the signal events come from the pair-production of heavy parity-odd particles, and then they go through cascade decays which end at the dark matter particle. 
We will first consider all those $4$ jets coming from heavy parity-odd particle cascade decays and defer the cases with one jet from ISR in a later section. 

\begin{figure}[!ht]
\begin{center}
\includegraphics[width=0.5\textwidth]{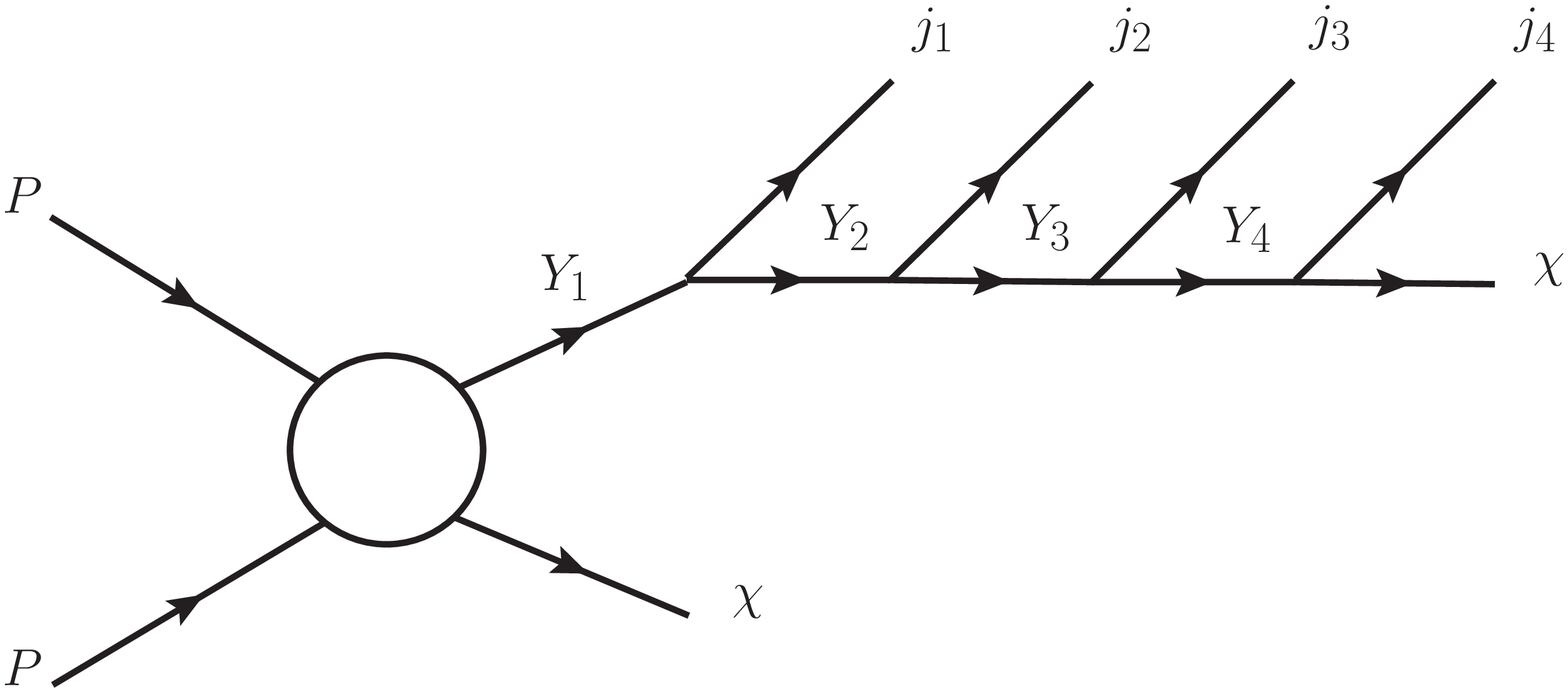} \hspace{0.5cm} \vspace{0.0cm}
\includegraphics[width=0.45\textwidth]{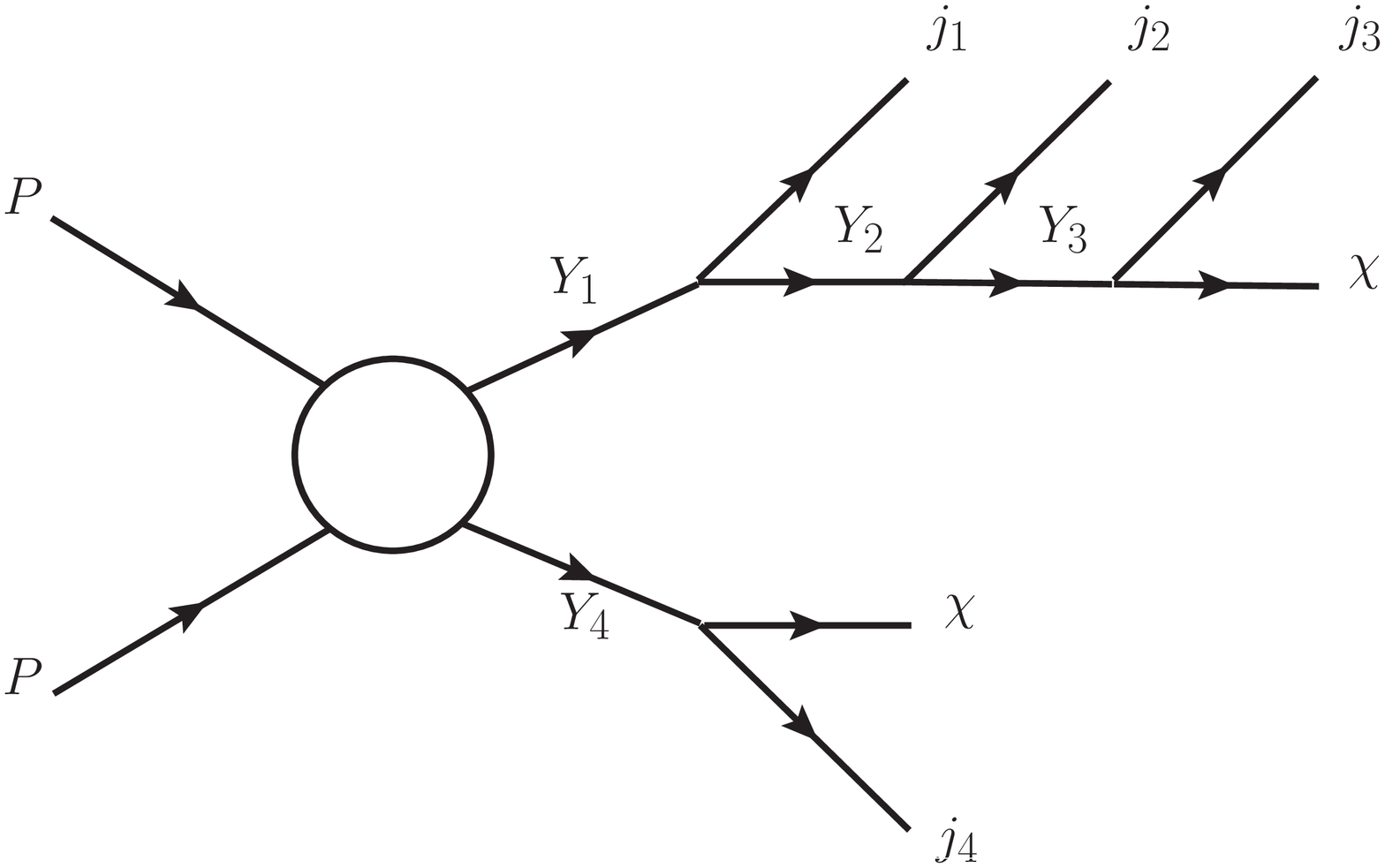}\\ \vspace{0.2cm}
\includegraphics[width=0.4\textwidth]{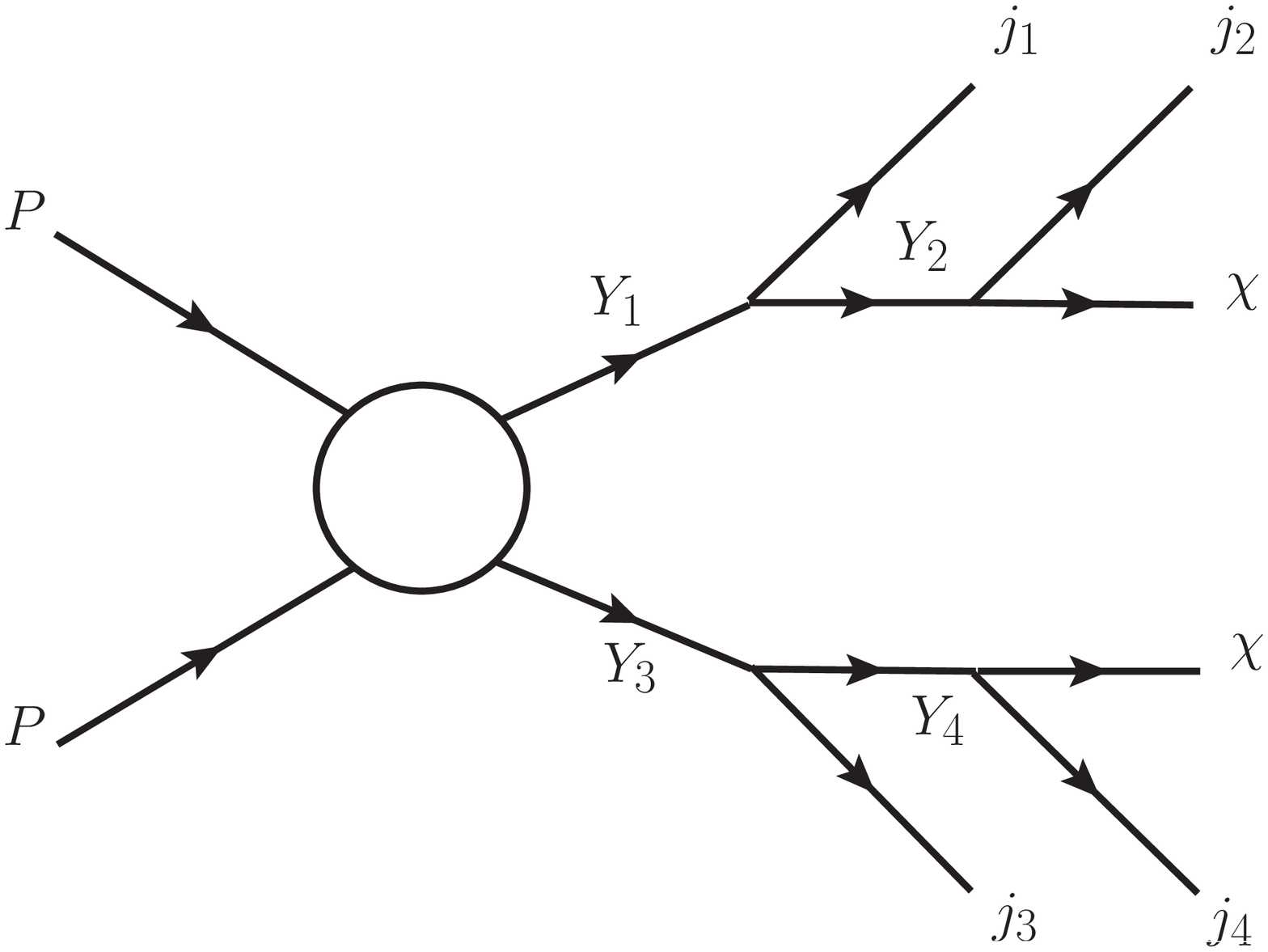}\\
\caption{The three event topologies, $\fourzero$, $\threeone$ and $\twotwo$, for $4j + \missET$ without specifying the spins of particles. The production mechanisms are not specified here and represented by a circle. The dark matter particle is denoted by $\chi$. The intermediated particles in each chain can be either on-shell or off-shell. 
}
\label{fig:feynfourjets}
\end{center}
\end{figure}

Under these assumptions, there are three topologies for this final state: four jets from one chain and zero jet from the other chain; three jets from one chain and one jet from the other chain; and two jets from both chains. The Feynman diagrams for those three cases are depicted in Fig.~\ref{fig:feynfourjets}.
%%%%
%%%%
In these diagrams, we denote the dark matter particle as $\chi$ and the decaying particles in the cascading chains as $Y_i$. The intermediate particles in each decay chain can be either on-shell or off-shell, both of which will be covered in the following sections. 

Although we do not restrict ourselves to any particular model, we note that all three topologies can arise in the popular SUSY scenario. In SUSY models, the lightest neutralino is usually the lightest supersymmetric particle (LSP) and also the dark matter particle. The $\fourzero$ case can come from the associated production of the lightest neutralino and gluino with the gluino going through the cascade decays: $pp \rightarrow \tilde{g} + \chi$ with $\tilde{g} \rightarrow \tilde{u}_L + 1j \rightarrow \chi_2 + 2j \rightarrow \tilde{u}_R + 3j \rightarrow \chi + 4j$. The $\threeone$ case can arise from the squark pair production as $pp \rightarrow \tilde{u}_R + \tilde{u}_L$ with two squarks decaying  as $\tilde{u}_R \rightarrow \chi + j$ and $\tilde{u}_L \rightarrow \chi_2 + 1j \rightarrow \tilde{u}_R + 2j \rightarrow \chi + 3j$. Finally, the $\twotwo$ case can come from the gluino pair production, $pp \rightarrow \tilde{g} + \tilde{g}$, with the same decaying processes on both chains as $\tilde{g} \rightarrow \tilde{u}_R + 1j \rightarrow \chi + 2j$. Whether any of these processes occur and dominate the $4j + \missET$ signal events depends on the spectrum and details of the specific model. It is also possible that more than one processes give comparable contributions to  the particular final state. In this section we are looking for functions which can identify the event topologies, so we will consider the idealized case where all events come from a single process, and leave the more complicated situations to Section~\ref{sec:mix} after we discuss the realistic implementation of the strategy.

For the illustration purpose, in the following when we define our kinematic functions, we demonstrate them with a SUSY spectrum such that the LSP mass is $m_\chi = 200$~GeV and in each decay the mother parity-odd particle is heavier than the neighboring daughter parity-odd particle by 200~GeV in the decay chains for each topology. The jets coming from these on-shell decays will have similar energies and hence are indistinguishable. This is the most challenging scenario. If instead there are large hierarchies among these four jets, we can consider separately the invariant mass combinations of the jets based on their energy hierarchies and obtain more handles on whether the harder jets and/or softer jets come from the same decay chain. The partonic event simulations are generated with the Madgraph/MadEvents  \cite{Alwall:2007st} package for the 14 TeV center of mass energy of LHC with the CTEQ 6L1 \cite{Pumplin:2002vw} parton distribution functions.

The kinematic functions that we are looking for need to be able to distinguish different event topologies. As mentioned in the Introduction, the invariant mass combinations of visible particles coming from the same decay chain are constrained by the mass of the decaying mother particle and hence have end points in their distributions, while the invariant mass combinations of particles from different decay chains are not expected to exhibit the end-point structure. Therefore we focus on functions of various invariant mass combinations of the 4 jets. The end-point formulae for various invariant mass combinations of particles from the same decay chain are given in Appendix~\ref{sec:endpoint}. We found that the invariant mass distributions are sufficient and fast enough to achieve our goal. It might be worth exploring more complicated strategies and other kinematic variables to improve the results.

\begin{figure}[!ht]
\begin{center}
\includegraphics[width=0.5\textwidth]{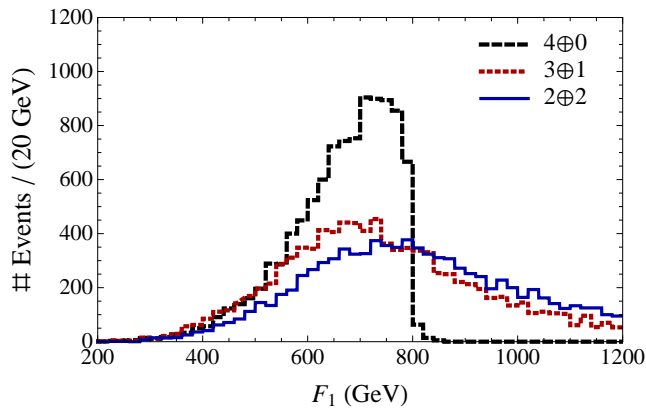} 
\caption{The number of events in terms of the function $F_1$ defined in Eq.~(\ref{eq:F1}) at the parton level. There are totally 10000 events for each topology generated in this plot. 
}
\label{fig:F1parton}
\end{center}
\end{figure}

To identify the $\fourzero$ topology, the obvious function to use is the total invariant mass distribution of all four jets.  We should anticipate a sharp end point for the $\fourzero$ case but not for the other two cases. Therefore, we define the first function, which is specifically sensitive to the $\fourzero$ topology, as
\beqa
F_1 (p_1, p_2, p_3, p_4) = \mbox{inv}[ p_1, p_2, p_3, p_4]\,.
\label{eq:F1}
\eeqa
Here, $p_i$ is defined according to the ordering of jet $E_T$ in each event and $\mbox{inv}[\,,\cdots\,,\,]$ means the total invariant mass of all momenta in the bracket, $\sqrt{(\sum_i p_i)^2}$. The $F_1$ distributions for three different topologies are shown in Fig.~\ref{fig:F1parton}.
%%%%
%%%%
One can see a sharp end-point at $800$~GeV for the $\fourzero$ topology in Fig.~\ref{fig:F1parton}. The sharp end point is expected to get smeared after showering, hadroniztion, and detector resolution effects. The strategy to deal with the realistic situations will be discussed in the next section.

\begin{figure}[!ht]
\begin{center}
\includegraphics[width=0.45\textwidth]{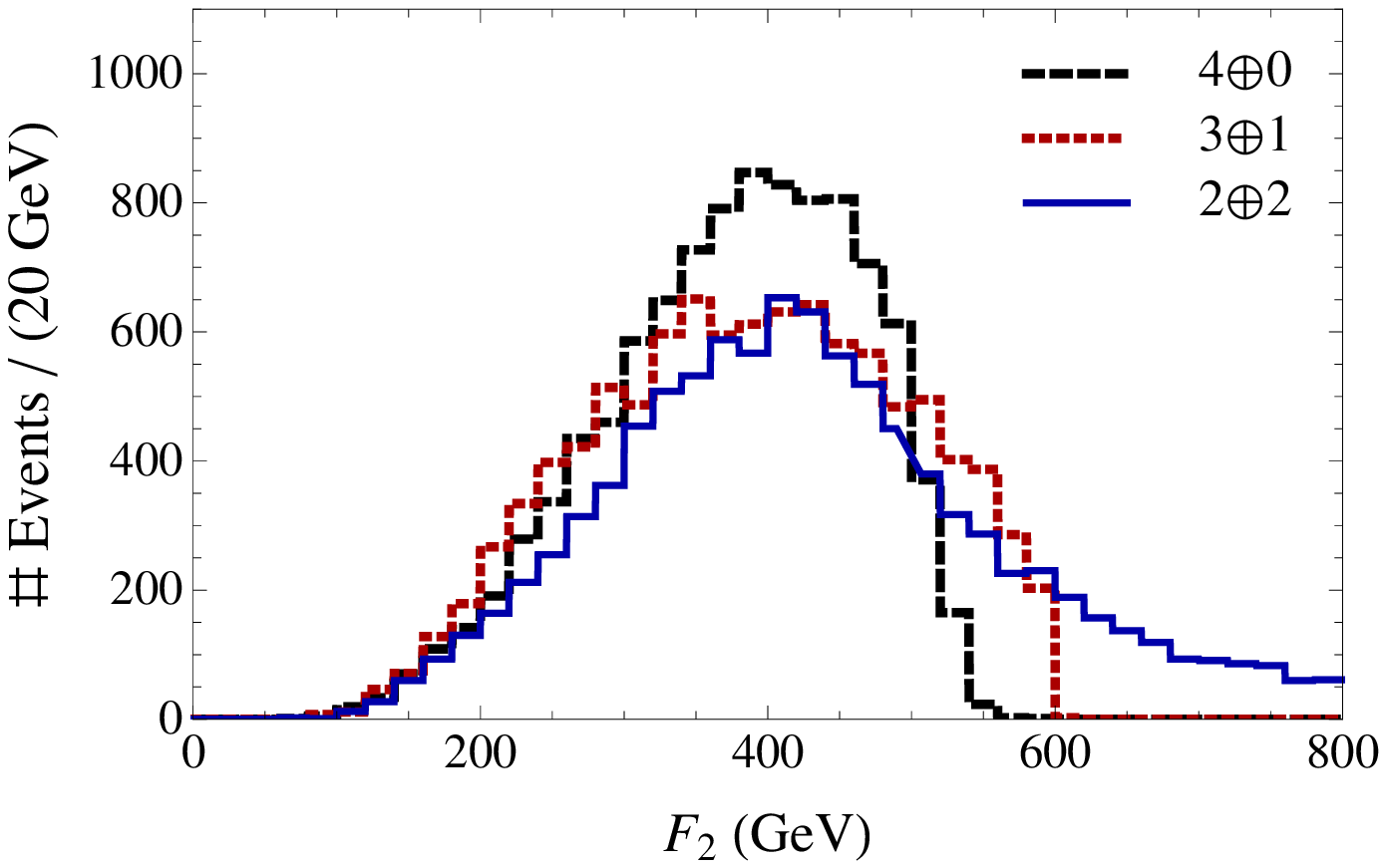}  \hspace{8mm}
\includegraphics[width=0.45\textwidth]{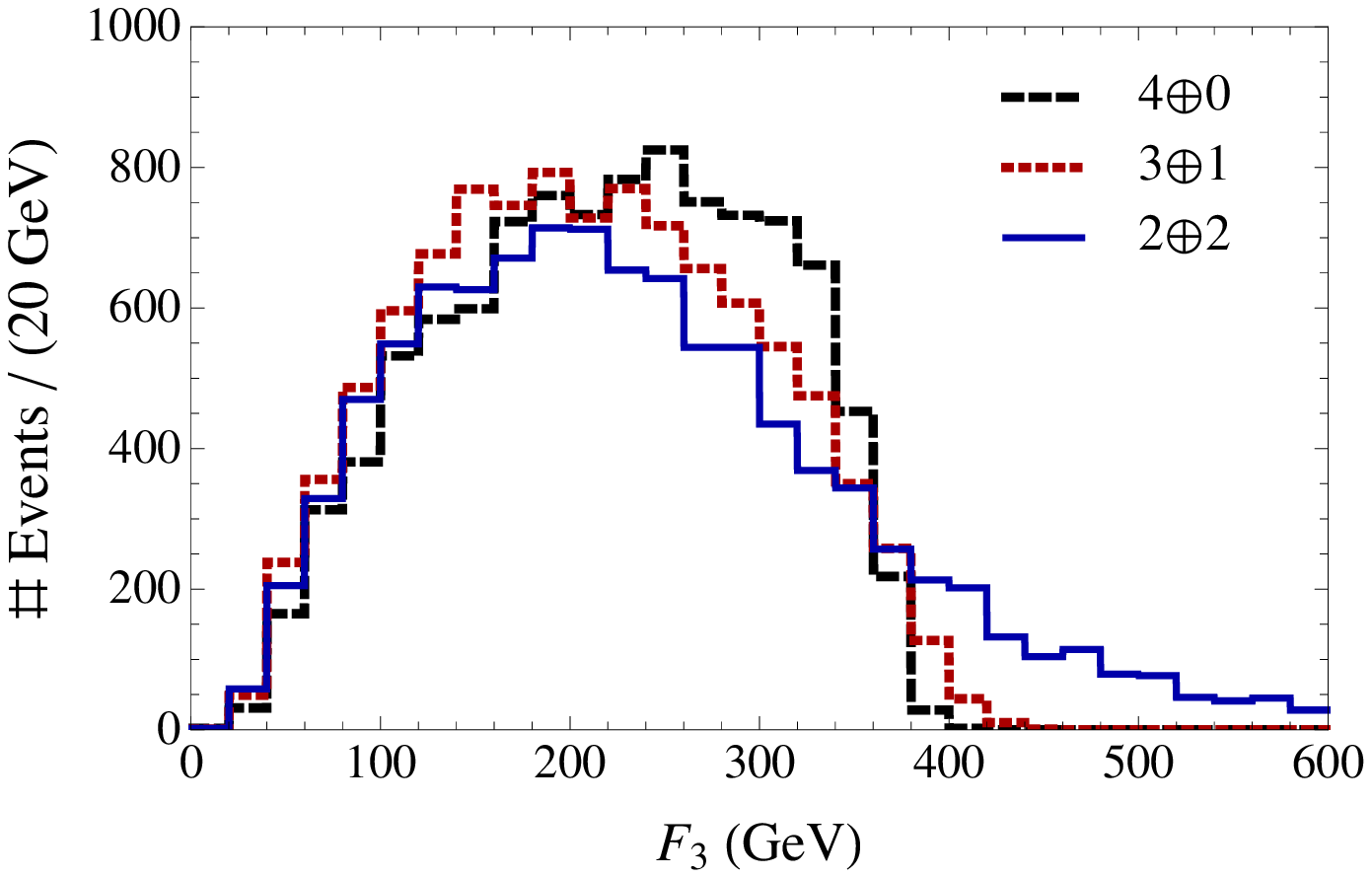} 
\caption{Left panel: the number of events in terms of the function $F_2$ at the parton level. There are totally 10000 events generated for each topology in this plot. Right panel: the same as the left one but in terms of $F_3$.
}
\label{fig:F2F3parton}
\end{center}
\end{figure}

For the $\threeone$ topology, we have found two potentially useful functions. The first one is the smallest 3-particle invariant mass combination:
\beqa
F_2(p_1, p_2, p_3, p_4) = \mbox{min} \big\{ \mbox{inv}[p_1, p_2, p_3], \mbox{inv}[p_1, p_2, p_4], \mbox{inv}[p_1, p_3, p_4] ,\mbox{inv}[p_2, p_3, p_4]  \big\} \,.
\label{eq:F2}
\eeqa
The minimum of the invariant masses of the four combinations has a high probability to find the set of three jets on the same chain. Even if it sometimes picks the wrong set, it will not exceed the expected end point since it is smaller than the correct combination. Plotting the event numbers as functions of $F_2$, one may expect a sharp end point for the $\threeone$ case but not for the $\twotwo$ case. An end point for the $\fourzero$ topology is also expected since all visible particles come from the same chain. Another function uses 2-particle invariant masses. Since the invariant mass of 2 particles from different chains can be very large, we consider the invariant mass of the pair of the particles which is opposite to the pair that forms the largest invariant mass:
\bea
F_3(p_1, p_2, p_3, p_4) &=& \mbox{inv}[p_k, p_l]  \,,\nonumber  \\
\mbox{such that} && \epsilon^{klij}\neq 0 \; \mbox{and} \;\mbox{inv}[p_i, p_j] = \mbox{max}\,\left\{\bigcup_{m, n}  \mbox{inv}[p_m, p_n]  \right\} \,.
\label{eq:F3}
\eea
Here, $i, j, k, l, m, n = 1, \cdots 4$ and $\epsilon^{klij}$ is the totally antisymmetric tensor. For the $\threeone$ case, this pair of particles has a large chance to come from the same chain. Even if occasionally they come from different chains, their invariant mass is bounded by an invariant mass from the same chain and hence will not exceed the corresponding end point. On the other hand, this combination for the $\twotwo$ case is likely to come from opposite chains and is not expected to have an end point. The event distributions of these two functions are shown in Fig.~\ref{fig:F2F3parton},
%%%%
%%%%
where one can see that both $\fourzero$ and $\threeone$ topologies have obvious end points at around $600$~GeV for $F_2$ and around 400~GeV for $F_3$. (The exact expected values can be found in Appendix~\ref{sec:endpoint}.) On the contrary, the $\twotwo$ distributions have long tails without end points. In the next section we find that $F_2$ seems to work better than $F_3$ after the experimental smearing. However, $F_3$ may still be useful for some other final states ({\it e.g.,} leptons which do not suffer too much from smearing effects).

\begin{figure}[!ht]
\begin{center}
\includegraphics[width=0.5\textwidth]{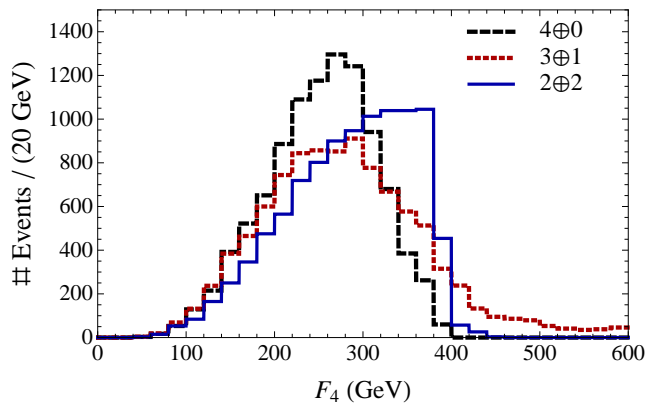} 
\caption{The number of events in terms of the function $F_4$ defined in Eq.~(\ref{eq:F4}) at the parton level. There are totally 10000 events generated for this plot. 
}
\label{fig:F4parton}
\end{center}
\end{figure}

To identify the $\twotwo$ topology, we define the following function which is sensitive to this topology:
\bea
F_4(p_1, p_2, p_3, p_4) &=& \mbox{min}\left\{   \bigcup_{i, j} \mbox{max} {\Big(} \mbox{inv}[i, j], \mbox{inv}[k, l] {\Big)} \right\} \quad \mbox{for} \quad \epsilon^{klij}\neq 0 \,.
\label{eq:F4}
\eea
For each event there are $3$ ways to pair those four jets. One first chooses the pair with the larger invariant mass for each way of pairing, then calculate the minimum of the larger invariant masses among those three partitions. For the $\twotwo$ case, the partition that each pair comes from the same decay chain will have both invariant masses of the two pairs to be bounded from above. Therefore, the correct partition is likely to have a smaller value for the maximum of the two invariant masses among the three partitions. On the other hand for the $\threeone$ topology, it is quite possible that the larger invariant mass pairs for  the three partitions all include the only jet from the shorter decay chain and therefore no end point is expected. The simulated results are shown in Fig.~\ref{fig:F4parton}.
%%%%
%%%%
Indeed one can see a sharp end point at around $400$~GeV for the $\twotwo$ topology but not for the $\threeone$ topology. The $\fourzero$ topology does have an end point but it is not very sharp.

At the parton level, one can see that the three different topologies can be easily identified by checking whether there are end points in those four functions $F_1$--$F_4$. However, including the showering, hadronization, and other experimental smearing effects these end points become less distinct and more concrete strategies need to be developed to identify the topologies, which is the subject of the next section. 

%%%%%%%%%%%%%%%%%%%%%%%%%%%%%%%%
\section{Dealing with Realistic Particle-Level Event Distributions}
\label{sec:particle}

In reality, the kinematic distributions proposed in the previous section will receive significant experimental smearing effects from showering, hadronization, detector resolutions, backgrounds, and so on. It is important to check whether the distinctive features of these functions can survive the smearing effects.  We also need an unambiguous  procedure to identify the event topology from these kinematic functions rather than just looking for end points by naked eyes. 

To include the experimental smearing effects for our particle-level analysis, we first generate parton-level events using Madgraph/MadEvents as before. We then process the parton level events with {\sc Pythia}~\cite{pythia} for showering and hadronization including initial and final state radiations, and PGS~\cite{pgs} (with the default CMS detector card) for the detector simulation. Some basic cuts are imposed on all signal events after the detector simulation.

We start with the same spectra used in the previous section with all on-shell decays and generate 10000 events for each topology. We require at least four jets with $E_T > 100$~GeV and the missing transverse energy $\missET > 200$~GeV on the events. This set of cuts are utilized just for the illustration purpose, and our following analysis is insensitive to those cuts. The signal selection efficiencies of $\fourzero$, $\threeone$ and $\twotwo$ are 19.3\%, 10.1\% and 13.1\%, respectively. 

\begin{figure}[!ht]
\begin{center}
\includegraphics[width=0.45\textwidth]{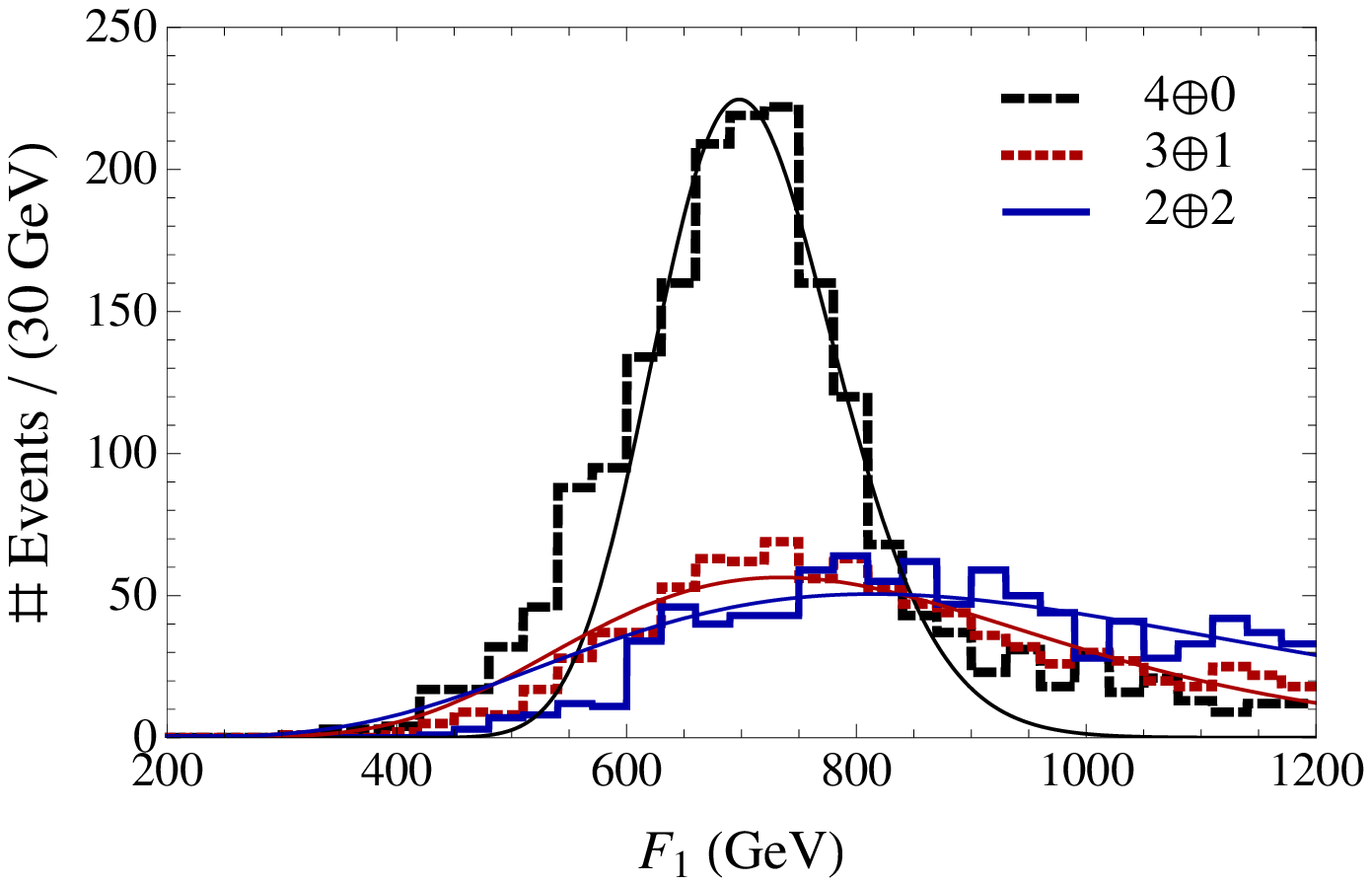}  \hspace{8mm}
\includegraphics[width=0.45\textwidth]{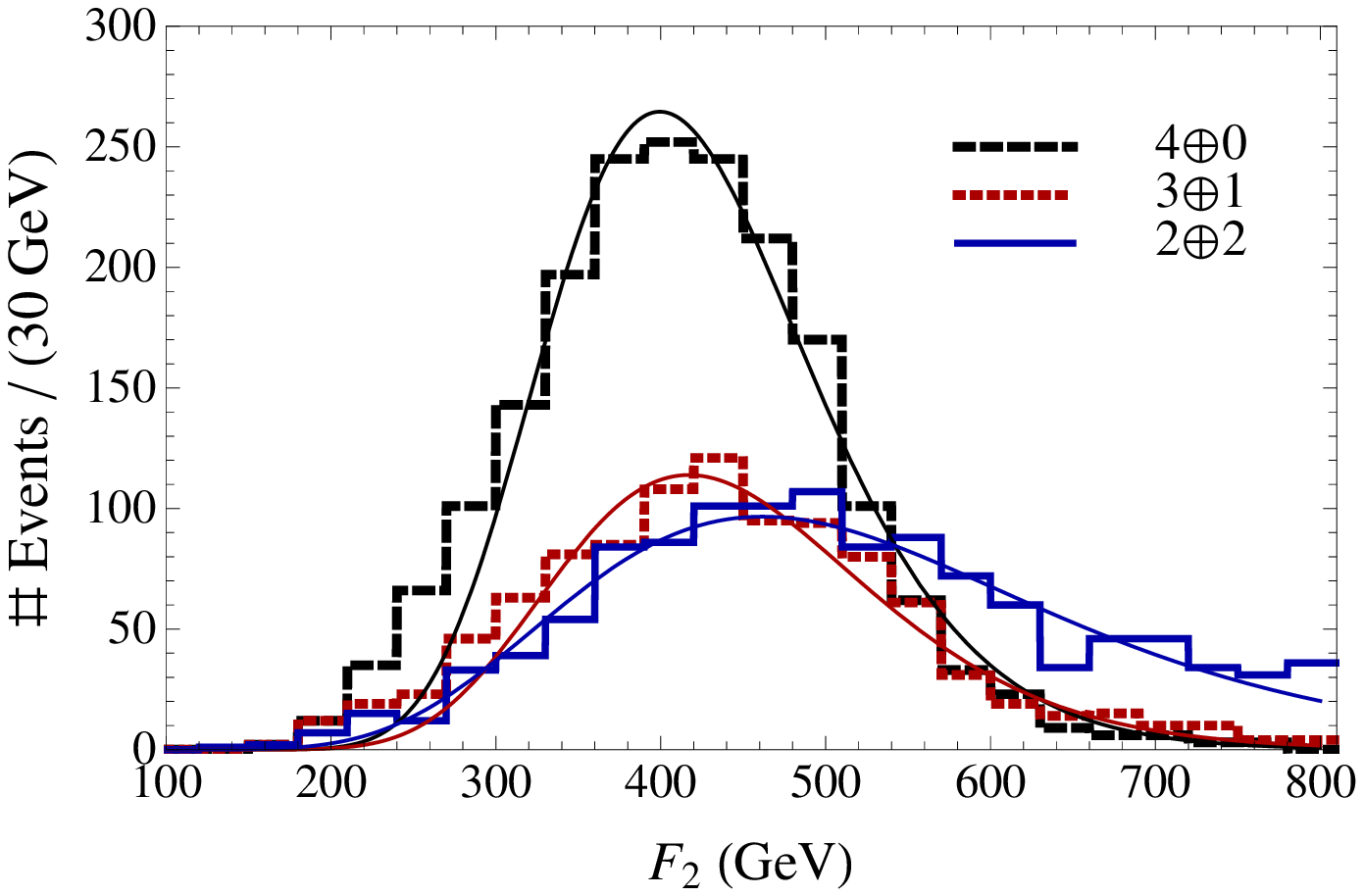}  \vspace{4mm} \\
\includegraphics[width=0.45\textwidth]{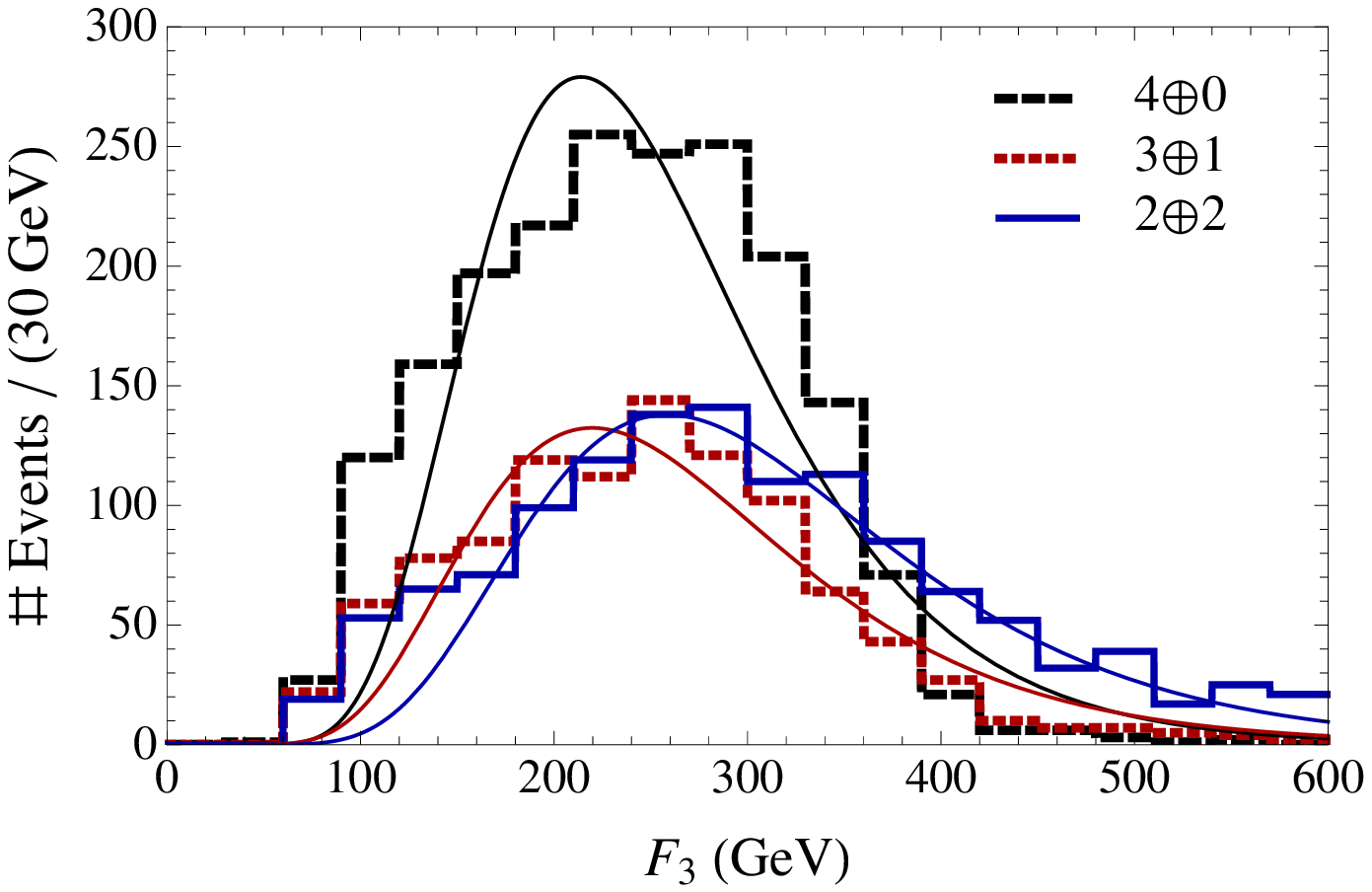}  \hspace{8mm}
\includegraphics[width=0.45\textwidth]{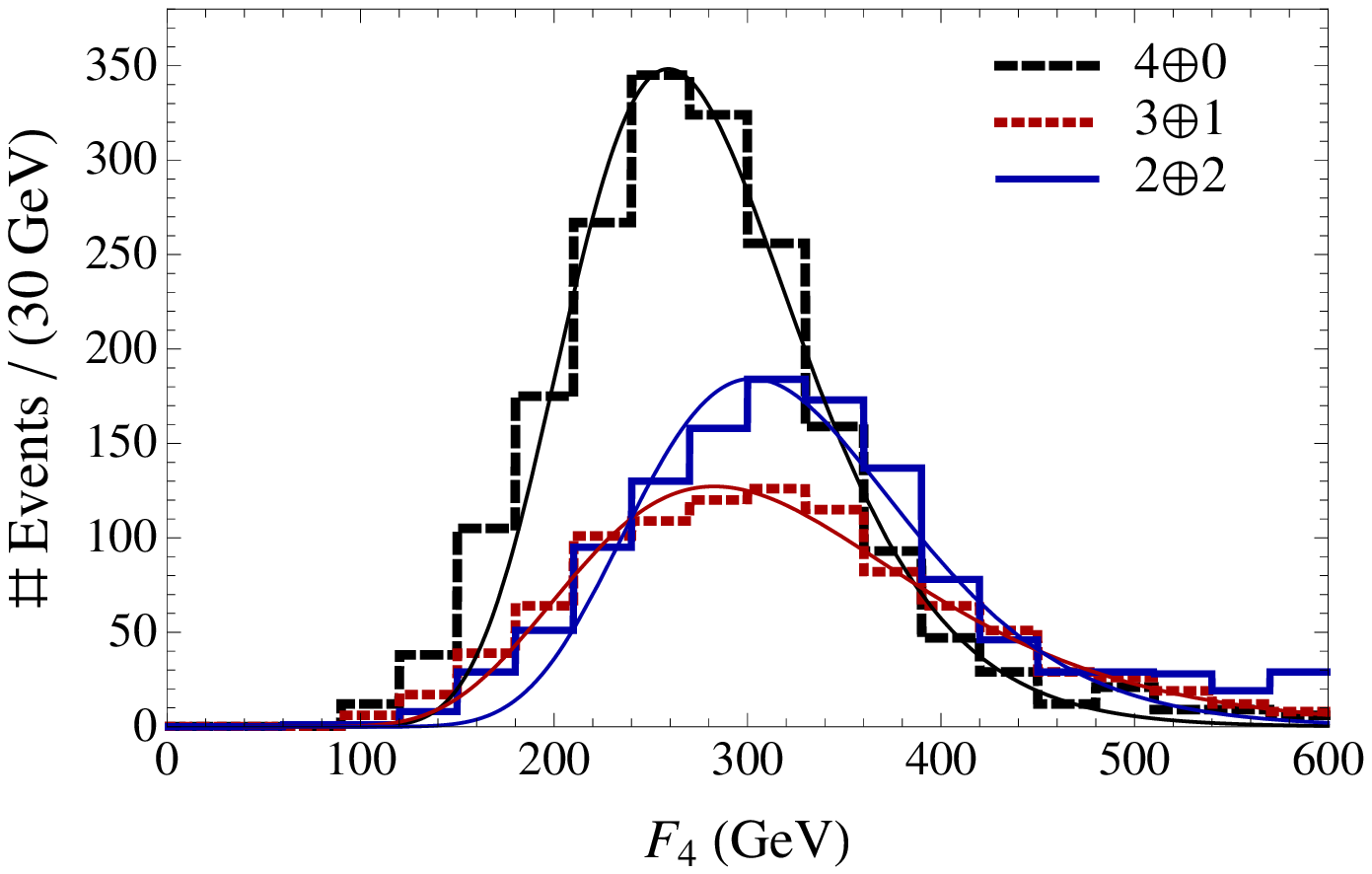} 
\caption{The number of events as functions of those four variables after passing {\sc Pythia}  and PGS. We select the signal events by requiring: at least 4 jets with $E_T > 100$~GeV and missing transverse energy $\missET > 200$~GeV. The continuous lines are fitted lines using the log-normal function. Only bins with a height above 1/2 of the peak height on the left side and above 1/4 of the peak height on the right side are included in the fit. 
}
\label{fig:Fallparticle-log-normal}
\end{center}
\end{figure}

The event distributions in terms of those four functions $F_i$ with $i = 1, \cdots\, 4$ are shown in Fig.~\ref{fig:Fallparticle-log-normal}. Compared to the parton level distributions in Fig.~\ref{fig:F1parton},~\ref{fig:F2F3parton},~\ref{fig:F4parton}, we can still roughly tell some of the end points for the first three functions. The end point structure for the last function $F_4$ is not so clear on the other hand. Nevertheless, for the distributions which were supposed to have end points, the after-peak slopes look steeper than the ones without end points. We will explore this observation to come up with a concrete procedure to identify different topologies.
%%%%
%%%%

To find the falling slopes after the peaks of these distributions, we will try to fit the histograms in Fig.~\ref{fig:Fallparticle-log-normal} with some simple functions. The first fitting function that we consider is the following log-normal function:
\beq
h_1 (F_i) \,=\, a_0\, \times\, \exp\left\{- \left(\frac{\ln[F_i/a_1]}{a_2} \right)^2\right\} \,.
\label{eq:log-normal}
\eeq
This log-normal function has a peak structure with asymmetric half-height widths. The three parameters $a_0$, $a_1$ and $a_2$ determine the peak height, the peak location and the falling slope after the peak. The last parameter $a_2$ can be used to distinguish different topologies. A smaller value of $a_2$ means a steeper slope after the peak. Given the large statistical uncertainties on the tails of those distributions and also potentially large contaminations from backgrounds, we only include bins with a height above one half of the peak height on the left side and above one quarter of the peak height on the right side into the fit. The $\chi^2$ in our fit is defined as
\bea
\chi^2 \,=\, \sum_{i_{\rm bin}} \frac{ (h_{1}^{i_{\rm bin}} - s^{i_{\rm bin}})^2   }{s^{i_{\rm bin}}} \,,
\eea
to only take into account the statistical uncertainties from the signal events. The asymmetric choice of bins around the peak is to minimize the effects of cuts, which affect the left side of the peak more severely. The fitted curves are shown in the continuous lines in Fig.~\ref{fig:Fallparticle-log-normal} for the 30 GeV bin size. As one can see, except the $F_3$ distributions all other histograms are well fitted by the log-normal functions.

The fitted values of $a_2$ are listed in Table~\ref{tab:a2fit30GeV} for 30 GeV and 40 GeV bin sizes.
\begin{table}[ht!]
\vspace*{4mm}
\renewcommand{\arraystretch}{1.5}
\centerline{
\begin{tabular}{|c||cccc||cccc|}
\hline
 & \multicolumn{4}{c||}{bin size = 30 GeV} & \multicolumn{4}{c|}{bin size = 40 GeV}  \\ \hline
   Topologies       & $a_2(F_1)$ & $a_2(F_2)$ & $a_2(F_3)$ & $a_2(F_4)$   
      & $a_2(F_1)$ & $a_2(F_2)$ & $a_2(F_3)$ & $a_2(F_4)$ \\ \hline
$\fourzero$  & 0.16 & 0.29  & 0.48 & 0.32   
 & 0.16 & 0.29  & 0.52 & 0.30         \\ \hline
$\threeone$ & 0.40  & 0.32  & 0.53 & 0.43   
 & 0.39  & 0.30  & 0.51 & 0.43 \\ \hline 
$\twotwo$   & 0.52  & 0.44  & 0.52 & 0.32    
   & 0.52  & 0.48  & 0.51 & 0.29         \\ \hline \hline
\end{tabular}
}
\caption{The fitted values for $a_2$ of the log-normal function in Eq.~(\ref{eq:log-normal}), which determine the steepness of the slopes of the histograms after the peak.  The bin size has been chosen to be 30~GeV (left), and 40~GeV (right). 
}
\label{tab:a2fit30GeV}
\end{table}
Smaller values of the bin size introduce larger fluctuations of the number of events in each bin, and make the fit unreliable. Comparing those numbers for the 30 GeV and the 40 GeV bin sizes, we can see that the fitted slopes are fairly consistent.  

\begin{figure}[ht!]
\begin{center}
\includegraphics[width=0.45\textwidth]{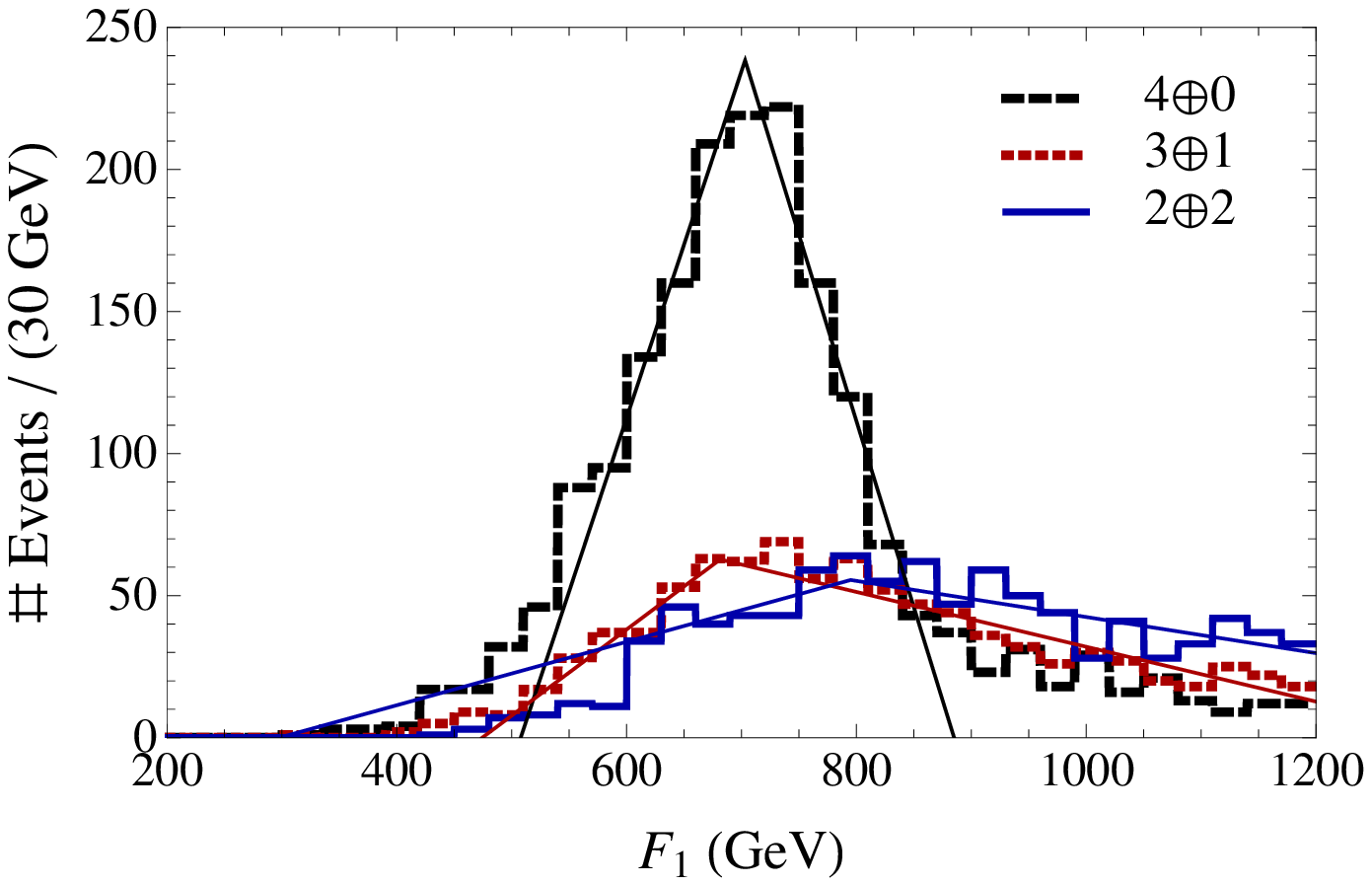}  \hspace{8mm}
\includegraphics[width=0.45\textwidth]{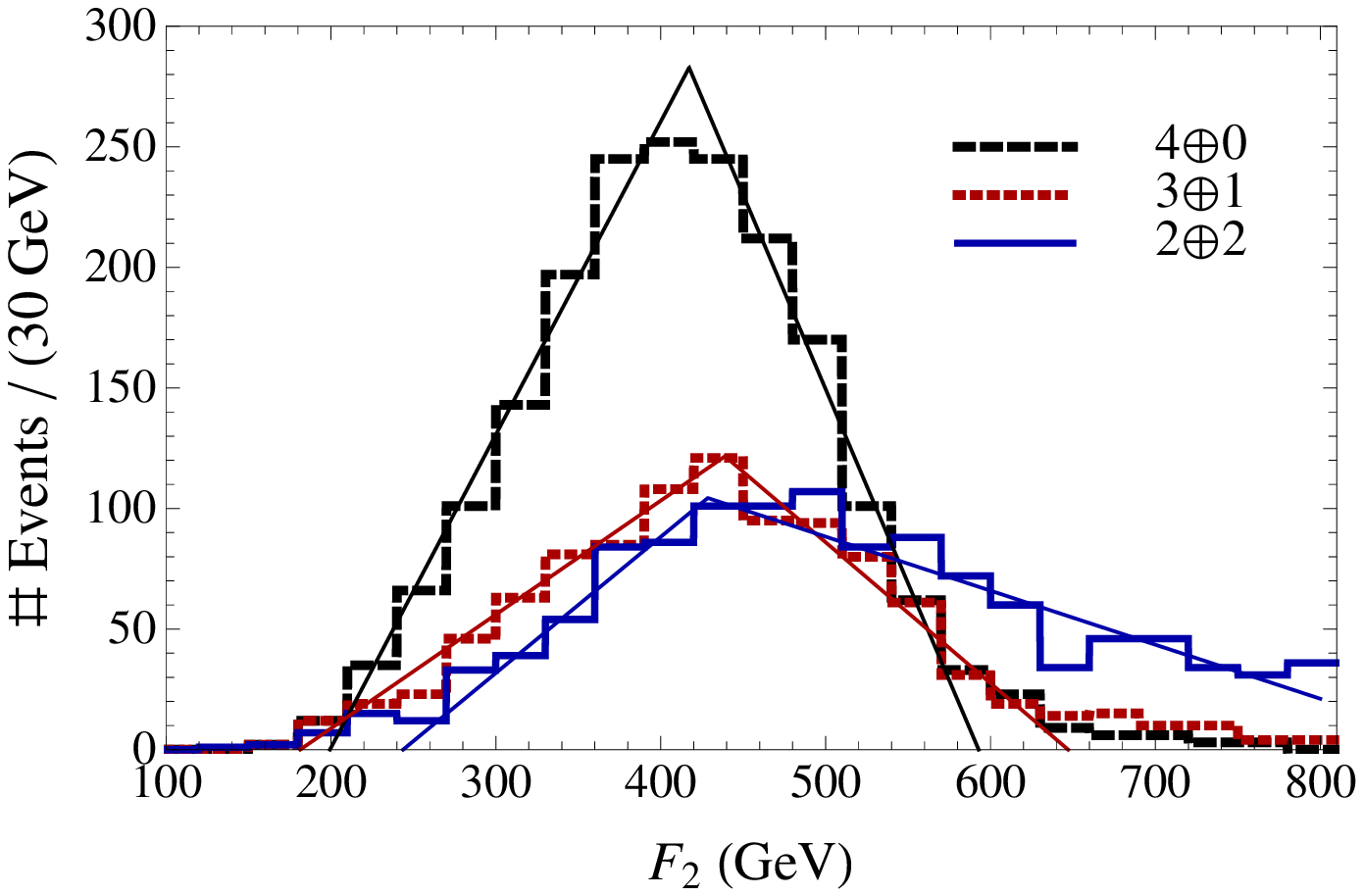}  \vspace{4mm} \\
\includegraphics[width=0.45\textwidth]{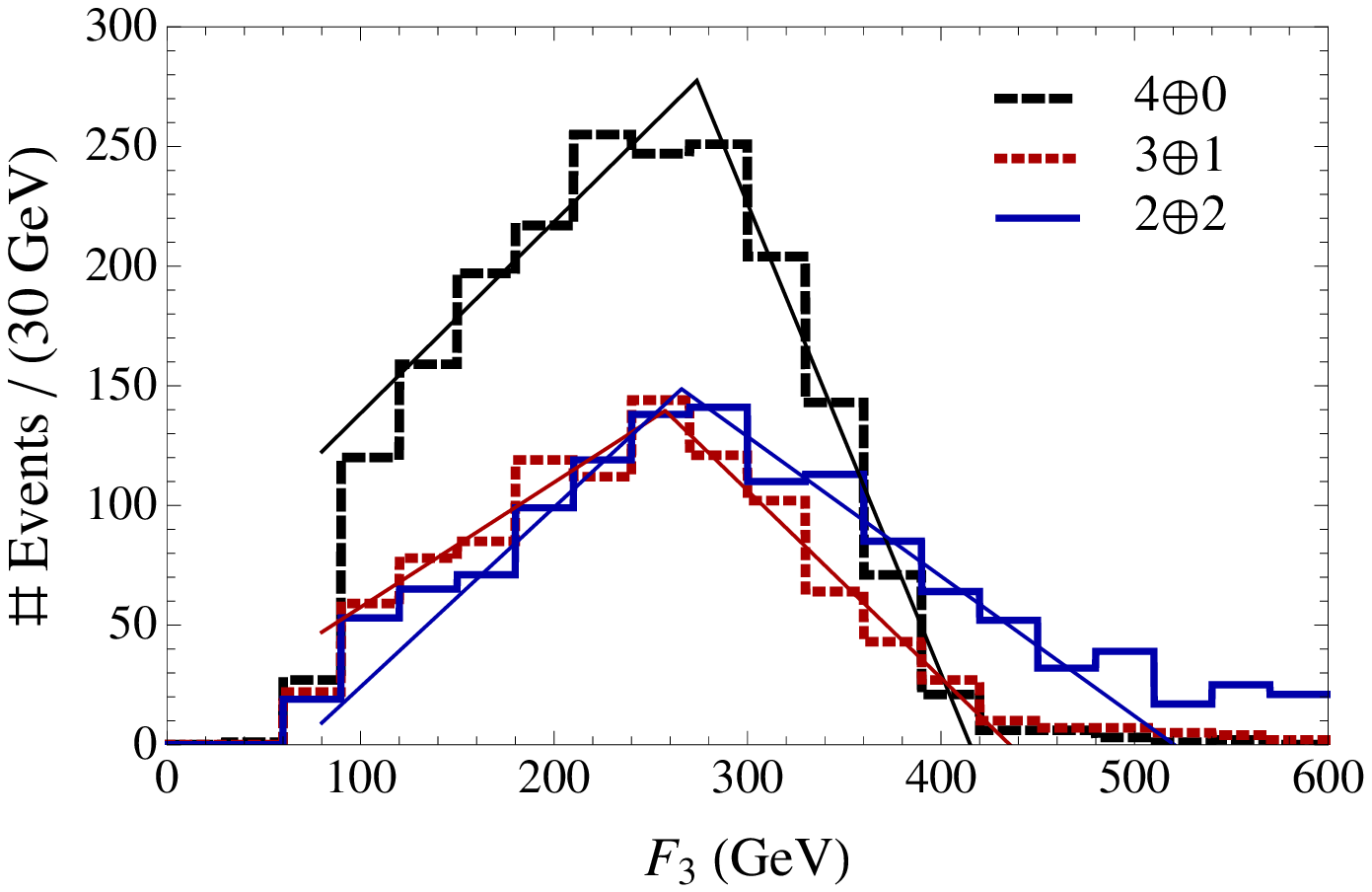}  \hspace{8mm}
\includegraphics[width=0.45\textwidth]{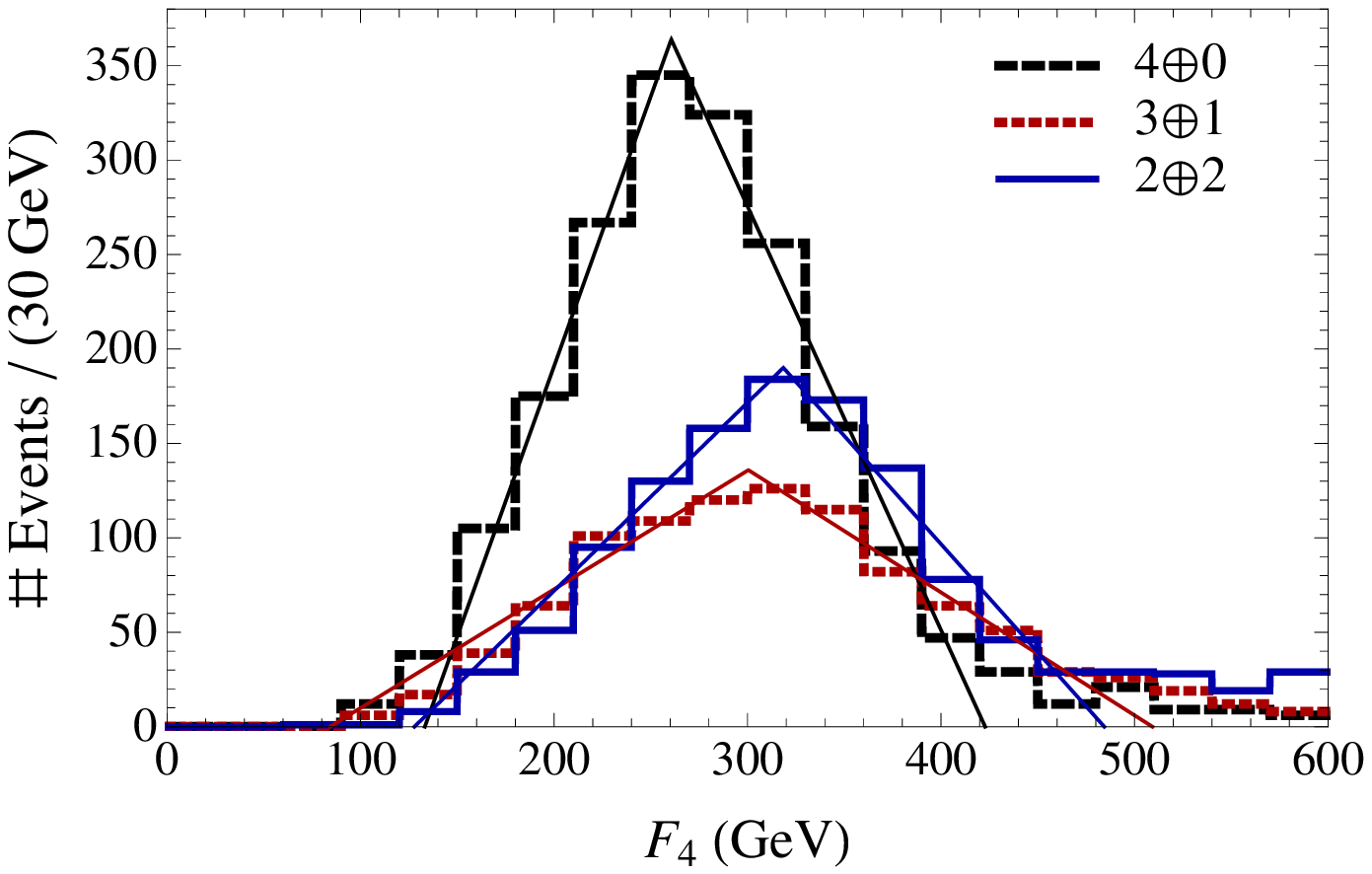} 
\caption{The same as Fig.~\ref{fig:Fallparticle-log-normal} but using the broken-line function defined in Eq.~(\ref{eq:broken-line}) to fit the distributions. 
}
\label{fig:Fallparticle-broken-line}
\end{center}
\end{figure}

To examine the sensitivity on the functions used to do the fits, we also try to fit the histograms with two straight lines around the peak. This fit also gives an estimate of the end point value at the point when the right straight line crosses zero. The ``broken-line'' function used in the fits is given by
\beq
h_2 (F_i) \,=\,
{\Big\{}
\renewcommand{\arraystretch}{1.5}
\begin{tabular}{ll}
$-c\,(F_i - a) + b$ & \mbox{for} \; $F_i \ge a$ \,, \\ 
$d\,(F_i - a) + b$ & \mbox{for} \; $F_i < a$\,,
\end{tabular} 
\label{eq:broken-line}
\eeq
with all parameters being positive. The parameters ``$(a, b)$" determine the location of the peak and ``$c$" and ``$d$" determine the slopes of curves on the right side and the left side of the peak. For the 30 GeV bin size, we compare the fitted results with the simulated distributions in Fig.~\ref{fig:Fallparticle-broken-line}.
%%%%
%%%%
We see that the ``broken-line" function can fit all histograms including $F_3$ pretty well. The slope ``$c$'' depends on the number of events and the overall scale of the invariant mass functions. To quantify the differences of the slopes on the right side of the peaks for different topologies and functions, we define the dimensionless combination $b/(ac)$ to be inverse of the normalized slope, which is independent of the number of events and the overall scale. A smaller value of $b/(ac)$ means a steeper slope on the right side of the peak.  For two different bin sizes, 30 GeV and 40 GeV, we present the fitted values of $b/(ac)$ in Table~\ref{tab:bacfit30GeV}. 
\begin{table}[ht!]
\vspace*{4mm}
%%%%%
\renewcommand{\arraystretch}{1.3}
\centerline{
\begin{tabular}{|c||cccc||cccc|}
\hline
 & \multicolumn{4}{c||}{bin size = 30 GeV} & \multicolumn{4}{c|}{bin size = 40 GeV}  \\ \hline
   Topologies       & $\frac{b}{ac}(F_1)$ & $\frac{b}{ac}(F_2)$ & $\frac{b}{ac}(F_3)$ & $\frac{b}{ac}(F_4)$   
      & $\frac{b}{ac}(F_1)$ & $\frac{b}{ac}(F_2)$ & $\frac{b}{ac}(F_3)$ & $\frac{b}{ac}(F_4)$ \\ \hline
$\fourzero$  & $0.26\pm 0.02$ & $0.42\pm 0.04$  & $0.52\pm 0.03$ & $0.62\pm 0.04$   
 & 0.24 & 0.48  & 0.47 & 0.49         \\ \hline
$\threeone$ & $0.95\pm 0.19$  & $0.47\pm 0.06$  & $0.69\pm 0.09$ & $0.70 \pm 0.10$  
 & 0.95  & 0.49  & 0.70 & 0.70 \\ \hline 
$\twotwo$   & $1.10\pm 0.30$  & $1.09\pm 0.15$  & $0.96\pm 0.14$ & $0.52\pm 0.05$    
   & 1.09  & 1.09  & 1.21 & 0.53         \\ \hline \hline
\end{tabular}
}
\caption{The fitted values for  $b/(ac)$ of the ``broken-line'' function in Eq.~(\ref{eq:broken-line}), which determine the normalized steepness of the slope of the histogram after the peak. The bin size has been chosen to be 30~GeV (left), and 40~GeV (right). We have also shown the $1\,\sigma$ statistical uncertainties for the 30 GeV bin size, based on 1932, 1013 and 1313 events after cuts for $\fourzero$, $\threeone$ and $\twotwo$, respectively.
}
\label{tab:bacfit30GeV}
\end{table}
Similarly, the fitted results are insensitive to the choices of bin sizes. 

One can na\"ively define the end point to be the intersecting point of the right branch of the broken-line with the $x$-axis, which is given by $F^{\rm end}_i \equiv a+b/c$, as a simple estimate of the true end point.  
\begin{table}[ht!]
\vspace*{4mm}
\renewcommand{\arraystretch}{1.0}
\centerline{
\begin{tabular}{|c|| llll  || llll |}
\hline
 & \multicolumn{4}{c||}{bin size = 30 GeV} & \multicolumn{4}{c|}{bin size = 40 GeV}  \\ \hline
       & $F^{\rm end}_1$ & $F^{\rm end}_2$ & $F^{\rm end}_3$ & $F^{\rm end}_4$   
      & $F^{\rm end}_1$ & $F^{\rm end}_2$ & $F^{\rm end}_3$ & $F^{\rm end}_4$ \\ \hline
$\fourzero$  & 885$\pm 19$ (800) & 592$\pm 19$ (600)  & 415$\pm 10$ (394) & 422$\pm 11$ (397)   
 & 886  & 603   & 408  & 409        \\ \hline
$\threeone$ & 1332  & 647$\pm 30$ (600)  & 436$\pm 25$ ($<458$) & 509   
 & 1329  & 652   & 441 & 512 \\ \hline 
$\twotwo$   & 1668  & 894  & 520 & 484$\pm 17$ (387)    
   & 1670  & 898  & 554 & 483        \\ \hline \hline
\end{tabular}
}
\caption{The fitted values (in GeV) of the end points, which are the right intersection point of broken lines with the $x$-axis. The bin size has been chosen to be 30~GeV (left), and 40~GeV (right). The numbers in the parenthesis are the end-point values at the parton level. We have also shown the $1\,\sigma$ statistical uncertainties for the same event samples as in Table~\ref{tab:bacfit30GeV} for the 30 GeV bin size.
}
\label{tab:endpoint}
\end{table}
The fitted values of the end points in terms of different functions are listed in Table~\ref{tab:endpoint}. As one might have expected, this definition tends to give a larger value of the end point than the actual value for most of the cases having a real end point.  

%%%%
\begin{figure}[ht!]
\begin{center}
\includegraphics[width=0.9\textwidth]{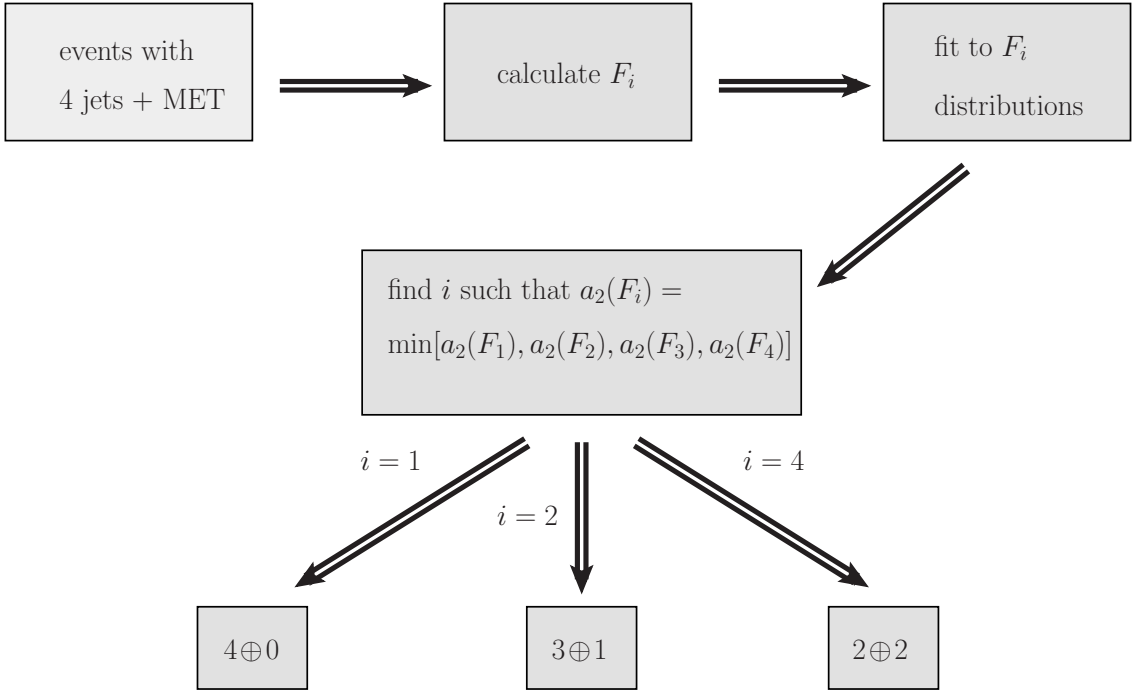} 
\caption{The flow chart  of our procedures to identify the event topologies with $4j + \missET$ when using the log-normal function to fit the $F_i$ distributions. For the broken-line fit, one needs to replace $a_2$ by $b/(ac)$. 
}
\label{fig:chartflow}
\end{center}
\end{figure}
%%%%

After fitting the shapes of distributions, we can now use the information in Table~\ref{tab:a2fit30GeV} and~\ref{tab:bacfit30GeV} to identify different topologies. From these tables, one can see that the topology can be identified by finding  ``$i$'' of the minimum of all four $a_2(F_i)$ for the log-normal fit or $b/(ac)(F_i)$ for the broken-line fit. If $i=1$, then the topology is $\fourzero$; if $i=2$ (or 3), it is $\threeone$; if $i=4$, it is $\twotwo$. The procedure described above to identify the dark matter event topologies with $4j+\missET$ is summarized in the flow chart  in Fig.~\ref{fig:chartflow}. The two choices of the fitting functions give similar results, so we will only use the broken-line fits in the rest of the paper.

Based on the central values and $1\,\sigma$ statistical uncertainties in Table~\ref{tab:bacfit30GeV}, we estimate the required numbers of events to achieve 90\% of time correct for the topology $\fourzero$ to be 446,  $\threeone$ to be 777 and $\twotwo$ to be 740. All of those numbers are defined after the basic kinematic cuts and assuming negligible SM backgrounds. A more complete analysis requires simulations of SM backgrounds and is beyond the scope of this paper. 

%%%%%%%%%%%%%%%%%%%%%%%%%%%%%%%%
\subsection{Initial State Radiation}
\label{sec:ISR}
There exist other possibilities to have the $4j+\missET$ final state after passing the basic kinematic cuts. For example, two of the jets come from a resonance like the $W$  boson. For this case, one can first check the two-body invariant mass distribution which should have an obvious mass peak, and then translate  the $4j+\missET$ final state to the $W + 2j +\missET$ final state. It could be easier to identify the three topologies for the $W + 2j +\missET$ final state following a similar method described before. Another possibility is to have the partonic events to be $5j +\missET$ but with one jet being soft and lost by basic cuts. If this indeed happens a lot, two dark matter parity-odd particles in the same decay chain may have masses close to each other. For this case, one can only treat those events as real $4j +\missET$ events and apply the procedure proposed earlier to at least understand the relative positions of the four hard jets, though the end points become less sharp and the formulae in Appendix~\ref{sec:endpoint} may not be accurate.

The third possibility is to have only 3 jets coming from the decay chains but the fourth jet from ISR. There are two types of topologies, $\threezeroISR$ (three jets from a single chain) and $\twooneISR$ (two jets from one chain and another jet from the other chain). In this subsection, we describe the additional procedures for identifying those two new topologies with one ISR jet. 

\begin{figure}[!ht]
\begin{center}
\includegraphics[width=0.5\textwidth]{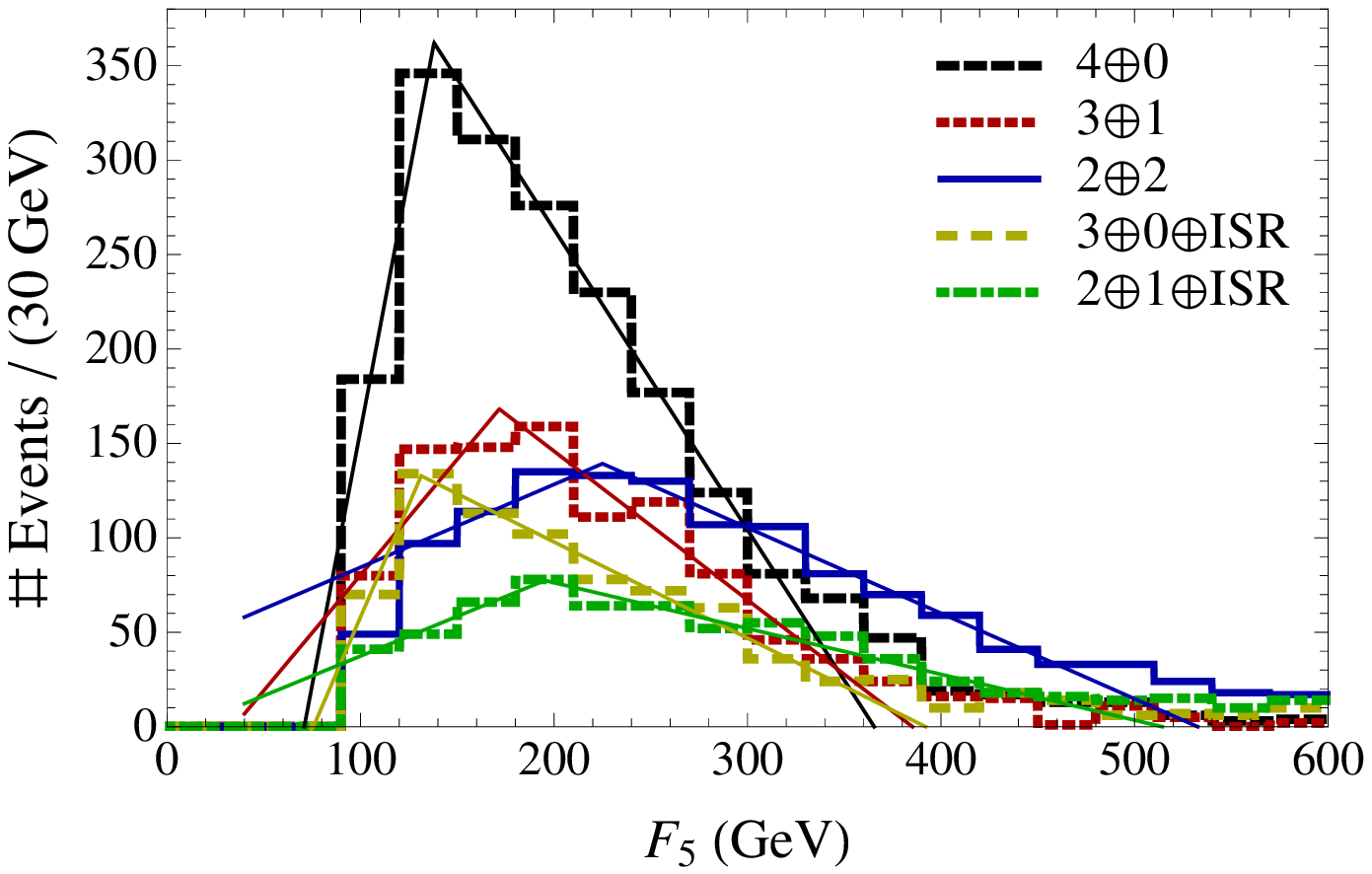} 
\caption{The distributions of $F_5$ for different topologies. 10000 events are generated for each topology before the cuts. There are 826 $\threezeroISR$ events and 699 $\twooneISR$ events after the cuts.
}
\label{fig:F5}
\end{center}
\end{figure}

The $\threezeroISR$ topology is very similar to the $\threeone$ topology. The only difference is that the isolated jet comes from ISR instead of from the decay of a heavy particle. Indeed, the fitted slopes for the $\threezeroISR$ topology have the same pattern as the $\threeone$ topology, with the $F_2$ function giving the steepest slope. To distinguish it from the $\threeone$ topology, we should examine the $E_T$ distribution of the isolated jet. If it comes from the ISR, it should have a falling $E_T$ distribution starting from the kinematic cut. On the other hand, a jet coming from a heavy particle decay should have a peak value determined by the mass difference. Because the 3 jets from the same decay chain are likely to have a smaller invariant mass, we define the following $\threezeroISR$-specified function:
\beqa
F_5(p_1, p_2, p_3, p_4) &=& E_T(p_k) \,,  \nonumber \\
\mbox{such that} &&
\epsilon^{klij}\neq 0 \; \mbox{and} \; \mbox{inv}[p_l, p_i, p_j] = \mbox{min} \big\{ 
\bigcup_{r, s, t} \mbox{inv}[p_r, p_s, p_t]
  \big\} \,.
\label{eq:F2}
\eeqa
The distributions of $F_5$ for different topologies are shown in Fig.~\ref{fig:F5}. 
%%%%
%%%%
There, we use the broken-line function to fit various distributions with 30 GeV bin size. As can be seen from this plot, $\threeone$ and $\threezeroISR$ have different locations of the peak with a lower value for the $\threezeroISR$ case. To quantify this difference, we compare the dimensionless parameter $d/c$, which measuring the ratio of the rising slope before the peak and the falling slope after the peak. The numerical values of the $b/(ac)$ and $d/c$  for those five topologies and five functions are listed in Table~\ref{tab:d/c30GeV}.
\begin{table}[ht!]
\vspace*{4mm}
\renewcommand{\arraystretch}{1.52}
%\centerline{
\begin{tabular}{|c||ccccc||ccccc|}
\hline
       & $\frac{b}{ac}(F_1)$ & $\frac{b}{ac}(F_2)$ & $\frac{b}{ac}(F_3)$ & $\frac{b}{ac}(F_4)$   & $\frac{b}{ac}(F_5)$ 
      & $\frac{d}{c}(F_1)$ & $\frac{d}{c}(F_2)$ & $\frac{d}{c}(F_3)$ & $\frac{d}{c}(F_4)$  & $\frac{d}{c}(F_5)$ \\ \hline \hline
$\fourzero$  & 0.26 & 0.42  & 0.52 & 0.62 & 1.65  
 & 0.93 & 0.81  & 0.41 & 1.27     & 3.39     \\ \hline
$\threeone$ & 0.95  & 0.47  & 0.69 & 0.70   & 1.24
 & 3.14  & 0.81  & 0.67 & 1.36 & 1.56 \\ \hline 
$\twotwo$   & 1.10  & 1.10  & 0.96 & 0.52    & 1.37 
   & 1.76  & 2.52  & 1.29 & 0.95    & 0.97     \\ \hline
$\threezeroISR$   & 0.84  & 0.48  & 0.60 & 0.93    & 1.98 
   & 1.51  & 0.98  & 0.46 & 2.39    & 4.68      \\ \hline   
$\twooneISR$   & 1.00  & 0.96  & 1.14 & 1.31    & 1.63 
   & 1.18  & 1.70  & 0.90 & 2.29    & 1.73      \\ \hline     
    \hline
\end{tabular}
%}
\caption{The fitted values of $b/(ac)$, which determine the normalized steepness of the slopes of the histograms after the peak, and the fitted values of $d/c$, which determine the ratios of the rising slopes and the falling slopes around the peak. The broken-line function in Eq.~(\ref{eq:broken-line}) is used to fit the distributions. The bin size has been chosen to be 30~GeV. 
}
\label{tab:d/c30GeV}
\end{table}
We see that the $\threezeroISR$ and $\threeone$ can be distinguished by examining $d/c(F_i)$. The $\threezeroISR$ has the largest value in $F_5$ while the largest value for $\threeone$ occurs at $F_1$. 
For the $\twooneISR$ topology, it is not expected to have an end point for any of these functions. Indeed, we find that all four $b/(ac)(F_i)$ are comparable and none of them takes a value as small as those ones with end points for the corresponding topology. That it is not any of the other topologies which should have at least one end point among the 4 functions can be taken as a sign for the $\twooneISR$ topology.  It is also not so easy to identify the ISR jet for this topology. To unambiguously distinguish it from other topologies is more challenging and probably requires additional functions. We will have more discussion on this topology in Appendix~{\ref{sec:twooneISR}}.

%%%%%%%%%%%%%%%%%%%%%%%%%%%%%%%%
\subsection{Off-Shell Decays}
\label{sec:Off-shell}

It frequently happens in models with dark matter that some particles have the dominant decays to be three-body processes.  For example in SUSY, if squarks are heavier than the gluino, the dominant decay channel of the gluino is to two quarks plus one neutralino via off-shell squarks. In this subsection, we use the SUSY model to generate events with off-shell decay processes in the decay chains with the assistance of the Monte-Carlo tool BRIDGE~\cite{Meade:2007js}. For the $\twotwo$ event topology, the events are generated by pair producing two gluinos, and then decaying each one to $\bar{u}+u+\chi$ via off-shell squarks. For $\threeone$, we first generate events with $\tilde{u}_L + \tilde{u}_R$ in the final state, and then we require $\tilde{u}_L \rightarrow \bar{u} + u+ \tilde{u}_R$ via the off-shell gluino or neutralino and $\tilde{u}_R \rightarrow u + \chi$. For $\fourzero$, we first generate events with $\tilde{g} + \chi$ in the final state, and then we require $\tilde{g} \rightarrow \bar{u} + u+ \chi_2$ and $\chi_2 \rightarrow \bar{u} + u + \chi$ via off-shell squarks. To produce events with similar visible kinematics for all three topologies, the LSP $\chi$ is fixed to have a mass of 200~GeV, and the mass difference between the mother superparticle and the daughter superparticle is chosen to be 200 GeV for two-body decays and 400 GeV for three-body decays. As a consequence, all four jets should have similar $E_T$ distributions on average. 

After the basic cuts  on all signal events: at least four jets with $E_T > 100$~GeV and the missing transverse energy $\missET > 200$~GeV, the acceptance efficiencies are $11.5\%$, $6.7\%$ and $9.2\%$ for $\fourzero$, $\threeone$ and $\twotwo$, respectively. Those efficiencies are lower than the on-shell decay cases simply because jets in the off-shell case have a larger probability to become soft. 

Repeating the same procedure as in the on-shell decay case, we have fitted the slopes of the distributions after the peak in Table~\ref{tab:b/acfitOffshell}. 
\begin{table}[ht!]
\vspace*{4mm}
\renewcommand{\arraystretch}{1.5}
\centerline{
\begin{tabular}{|c||cccc||cccc|}
\hline
 & \multicolumn{4}{c||}{bin size = 25 GeV} & \multicolumn{4}{c|}{bin size = 30 GeV}  \\ \hline
   Topologies       & $\frac{b}{ac}(F_1)$ & $\frac{b}{ac}(F_2)$ & $\frac{b}{ac}(F_3)$ & $\frac{b}{ac}(F_4)$   
      & $\frac{b}{ac}(F_1)$ & $\frac{b}{ac}(F_2)$ & $\frac{b}{ac}(F_3)$ & $\frac{b}{ac}(F_4)$ \\ \hline
$\fourzero$  & 0.40 & 0.93  & 0.92 & 0.82  
 & 0.44 & 0.85  & 0.82 & 0.78         \\ \hline
$\threeone$ & 0.96  & 0.62  & 0.80 & 0.80   
 & 0.80  & 0.67  & 0.58 & 0.84 \\ \hline 
$\twotwo$   & 1.47  & 1.09  & 0.84 & 0.74    
   & 1.47  & 1.15  & 0.83 & 0.81         \\ \hline \hline
\end{tabular}
%\label{tab:b/acfitOffshell}
}
\caption{The fitted values of $b/(ac)$ for the off-shell decay case. The broken-line fit is used to obtain those numbers. Again each topology has 10000 events before the cuts, and there are 1152, 671 and 915 events passing the cuts for $\fourzero$, $\threeone$ and $\twotwo$ respectively.
}
\label{tab:b/acfitOffshell}
\end{table}
Comparing Table~\ref{tab:b/acfitOffshell}  with Table~\ref{tab:bacfit30GeV} for the on-shell case, we can see that the differences of $b/(ac)(F_i)$ are reduced for the off-shell case. Although a similar selection criterion like Fig.~\ref{fig:chartflow} can still be used to identify those three topologies, more signal events or a higher  integrated luminosity at the LHC are required to identify the  topologies for the off-shell case. After the event topology is identified, the on-shell and off-shell decay cases may be distinguished by examining the correlations  of the invariant masses~\cite{Burns:2009zi} and/or the function $M_{T2,{\rm max}}$~\cite{Lester:1999tx,Barr:2003rg,Cho:2007qv,Barr:2007hy,Cho:2007dh}.

For completeness, we also report the fitted end-point values in Table~\ref{tab:endpoint-offshell} for the 30 GeV bin size.
\begin{table}[ht!]
\vspace*{4mm}
\renewcommand{\arraystretch}{1.5}
\centerline{
\begin{tabular}{|c|| llll  || }
\hline
 & \multicolumn{4}{c||}{bin size = 30 GeV}  \\ \hline
       & $F^{\rm end}_1$ & $F^{\rm end}_2$ & $F^{\rm end}_3$ & $F^{\rm end}_4$   
       \\ \hline
$\fourzero$  & 977 (800) & 660 ($<754^*$)  & 411 ($<754^*$) & 438 ($<754^*$)   
     \\ \hline
$\threeone$ & 1343  & 669 (600)  & 413 ($<600^*$) & 538  
  \\ \hline 
$\twotwo$   & 1680  & 806  & 485 & 483 (400)    
       \\ \hline \hline
\end{tabular}
}
\caption{The fitted values (in GeV) of the end points, which are the right intersection points of broken lines with the $x$-axis. The bin size has been chosen to be 30~GeV. The numbers in the parenthesis are the end-point values at the parton level. The numbers with $^*$ occur in the soft limit of some jet(s) which in practice will not pass the cuts. Therefore the actual end points are much smaller.
}
\label{tab:endpoint-offshell}
\end{table}
%

%%%%%%%%%%%%%%%%%%%%%%%%%%%%%%%%
\subsection{The Cases of Mixtures of Different Event Topologies}
\label{sec:mix}

\begin{figure}[!ht]
\begin{center}
\includegraphics[width=0.6\textwidth]{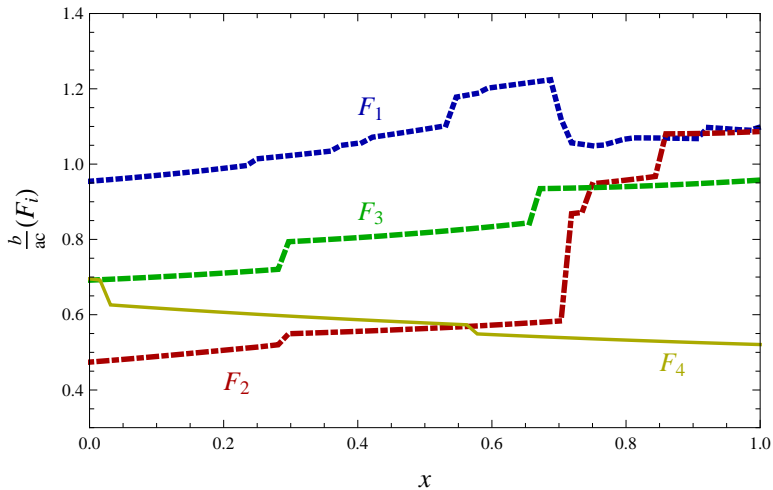} 
\caption{The fitted values of $b/(ac)$ for different functions $F_i$ as functions of $x$, where $x$ is the fraction of $\twotwo$ type events in a mixture of $\threeone$ and $\twotwo$ topologies with 1000 combined events (after the basic cuts). The bin size is 30 GeV. Only bins with a height above 1/2 of the peak height on the left side and above 1/4 of the peak height on the right side are included in the fit. 
}
\label{fig:mixture23}
\end{center}
\end{figure}

It may also happen that the signal events of the same final state come from a combination of two or more different event topologies. The method discussed in this paper should work when one topology dominates over the others. To quantify the limit at which the topology can be identified by our method, we use a mixture of $\twotwo$ and $\threeone$ topologies as a case study. Fixing the total number of events after the basic cuts to be 1000, we study the  patterns of the fitted $b/(ac)$ as a function of the mixture fraction, $x$. For $x=0$, all events are from the $\threeone$ topology, while for $x=1$ all events are from the $\twotwo$ topology. For the 4 functions $F_1$--$F_4$, the fitted values of $b/(ac)$ as functions of the mixture fraction are shown in Fig.~\ref{fig:mixture23}.
%%%%
%%%%
As $x$ increases, one can see from Fig.~\ref{fig:mixture23} that there is a transition of the function with the smallest $b/(ac)$ from $F_2$ to $F_4$. The transition happens around $x\approx 50\%$ with a mixture of equal amount of $\twotwo$ and $\threeone$ events. It is not surprising that if the signal is indeed a mixture of $\twotwo$ and $\threeone$ with comparable weights, one can not identify the topology unless additional kinematic information can split the signal events into the two categories. The abrupt changes of lines in Fig.~\ref{fig:mixture23} are due to changes of bins during the fit procedure by only including bins with a height above 1/2 of the peak height on the left side and above 1/4 of the peak height on the right side. 

\begin{figure}[!ht]
\begin{center}
\includegraphics[width=0.6\textwidth]{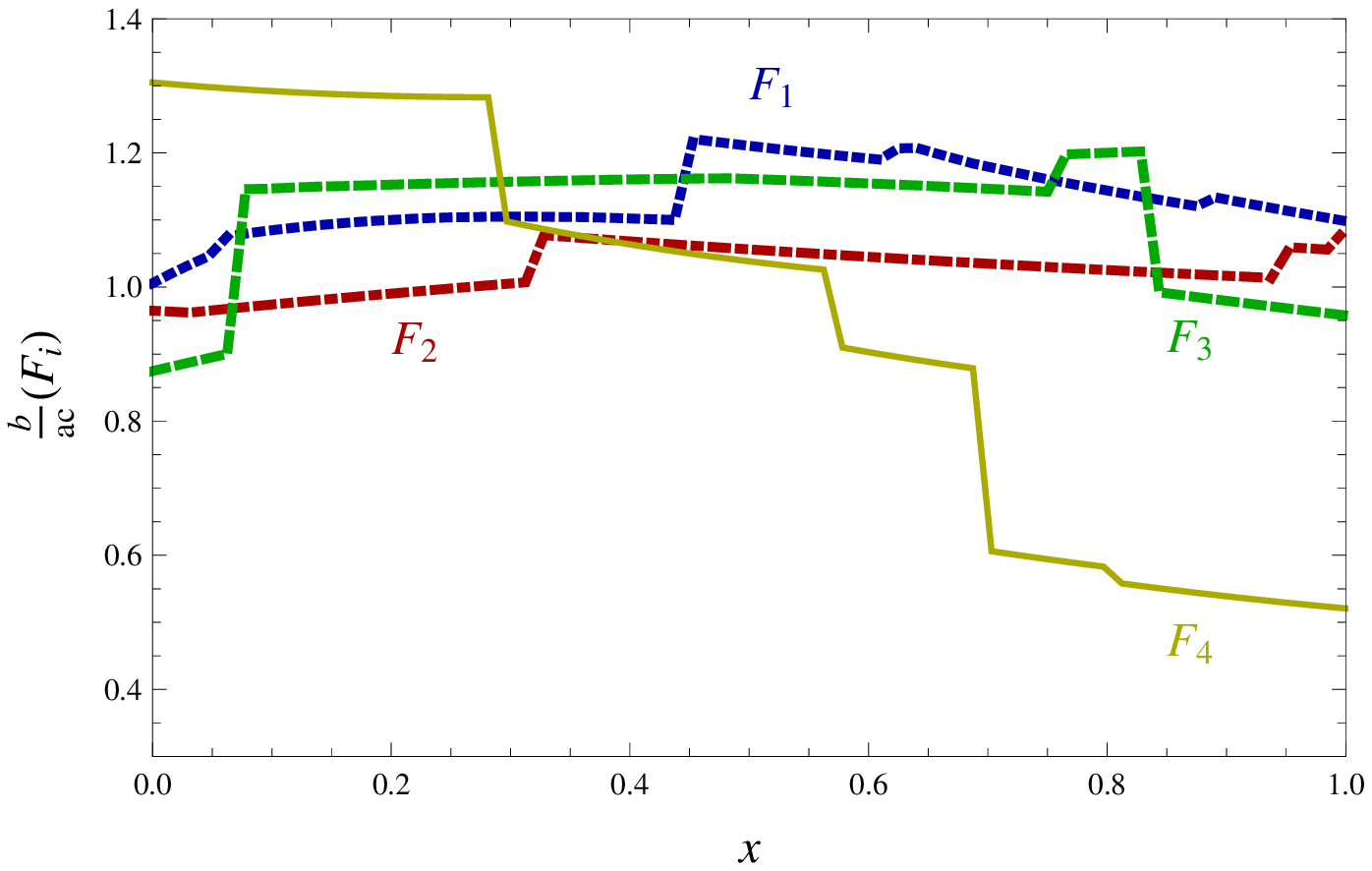} 
\caption{The same as Fig.~\ref{fig:mixture23} but for a mixture of $\twooneISR$ and $\twotwo$ topologies with 1000 combined events (after the cuts). Here, $x$ is the fraction of $\twotwo$ type events in a mixture of $\twooneISR$ and $\twotwo$ topologies.
}
\label{fig:mixture2ISR}
\end{center}
\end{figure}

In our analysis, the SM backgrounds are neglected. Since there is no reason to anticipate SM backgrounds to have end points in terms of those functions $F_i$, the pattern of the fitted slopes for different $F_i$ should not be modified much if the signal events dominate over the backgrounds. As an illustration, we treat the events from $\twooneISR$ as the backgrounds of $\twotwo$ signals, and show the patterns as a function of percentage of $\twotwo$ in Fig.~\ref{fig:mixture2ISR}. 
%%%%
%%%%
If the signal events are over around 50\% of the total number of events, the $b/(ac)(F_4)$ is the smallest among all $F_i$ functions. In this situation, we still can determine the signal events are from the $\twotwo$ topology.

%%%%%%%%%%%%%%%%%%%%%%%%%%%%%%%%
\section{Discussion and Conclusions}
\label{sec:Discussions}

Although we focus on the $4\,j + \missET$ final state in this paper, a similar set of functions can be used to identify event topologies of $n\,j + \missET$  with $n \geq 3$. Without ISR, there are $[n/2] + 1$  different topologies with two decay chains. For $n=2k+1$, one can define $k+1$ functions to find the minimum of all possible invariant masses of $m$ jets for $m = k+1, \cdots, 2k+1$. For $n = 2k$, other than the $k$ functions defined to be the minimum of all possible invariant masses of $m$ jets for $m = k+1, \cdots\, 2k+1$, one additional function can be defined as the minimum of the maximum of two invariant masses of $k$ jets among all possible combinations (similar to the function $F_4$). One can also define functions analogous to $F_3$ by removing some jets which form the largest invariant mass and looking at the the invariant mass of the other jets. For the cases with ISR, the general strategy would be to first identify the existence of an ISR jet by the jet $E_T$ distribution and then define functions based on the remaining $n-1$ jets. From our study of the $4\,j + \missET$ example, we found that some cases with ISR are difficult to identify without lowering the $E_T$ cuts on jets. For $2\,j +\missET$, the general strategy does not work. The case with 2 jets on the same chain may be checked by finding the end point of the 2-jet invariant mass. On the other hand, the topology with one jet on each decay chain can not be identified with invariant mass functions. Other kinematic functions like  $M_{T2}$ or  $M_{\rm CT}$~\cite{Tovey:2008ui} might be useful for this topology.

If the final states also contain leptons in addition to jets plus $\missET$, their charges, flavors and good energy resolutions can give better handles for the event topology identification. The lepton number and flavor conservations (if they are good symmetries) may already give us some hints on the type of topologies.  For instance, consider the $2\,j + 2\,\ell^+ + 2\,\ell^- + \missET$ final state. The SUSY-like theories suggest that the leptons from the same decay chain should have the same flavor and opposite charges. However, one can imagine more general models which may contain doubly charged particles and/or large lepton flavor violations and hence invalidate the argument. Nevertheless, the strategies taken in this paper should be able to identify event topologies without these assumptions. We can first take  the four leptons and use the methods in this paper to find out their distribution on the two chains. The case with 4 leptons on the same chain can be identified with the function $F_1$. The function $F_2$ (which is the minimum of the 3-lepton invariant masses) can be used to identify the case with 3 leptons from one chain and the other lepton from the other chain. The functions $F_3$ and $F_4$ can be divided into more functions based on whether we take the invariant mass of the 2 leptons of the same charge or the opposite charges. For example, the function $F_4$ restricted to opposite-charge invariant masses can identify the case with 2 opposite-charged leptons on each chain, while the pair of the same-charge 2-lepton invariant masses can identify the case with the same-sign leptons on each chain. Once the  relative distribution of the leptons on the 2 chains has been fixed, one can check the existence of the end point for the invariant mass distribution of two jets. If there is no end point, one simply assigns one jet on each chain. If there is an end point, the 2 jets are on the same chain. If the lepton distribution is not symmetric (for example, 3 leptons on one chain and one lepton on the other chain), one can then check the invariant mass combination of the 2 jets together with the 3 leptons to see if they come from the same chain. In this way, we know how the visible particles distributed between the two chains. To remove the order ambiguity of the jets and leptons on a single chain probably requires much more sophisticated analysis which may depend on the details of the event kinematic distributions.

In this paper we have generated events from SUSY models with certain specific spectra. The strategies employed in this paper should work for general dark matter models with an unbroken $\mathbb{Z}_2$ discrete symmetry~\cite{Cheng:2002iz,Cheng:2002ab,Cheng:2003ju,Cheng:2004yc,Low:2004xc,Bai:2010qg}, or even models with more complicated symmetries ({\it e.g.,} Ref.~\cite{Agashe:2010gt,Dulaney:2010dj,Batell:2010bp}) as long as the events contain 2 decay chains which end with missing particles. Additional missing particles like neutrinos coming from the decay chains may be difficult to be identified themselves, but should not prevent us from identifying the topology of the visible particle part. There could also be accompanied events with neutrinos replaced by its charged lepton partners which may be used to identify the topology. The invariant mass distributions may have different behaviors for different models. Depending on the spins of the intermediate particles, the sharpness of the end points of invariant mass distributions varies~\cite{Wang:2006hk}.  For example, comparing the processes of $\tilde{g} \rightarrow \bar{q} + \tilde{q} \rightarrow \bar{q} + q + \chi$ in SUSY and $g^{(1)} \rightarrow \bar{q} + q^{(1)} \rightarrow \bar{q} + q +B^{(1)}$ in the UED model, the invariant mass distribution of two jets in the UED model has a sharper end-point than in SUSY because it has a spin-1/2 intermediate particle, in contrast to the spin-0 particle in the SUSY case. In addition, the production cross section can vary a lot among different models with different spectra. Therefore, the actual required luminosity at the LHC to identify a particular topology depends on the specific model and requires a detailed study for the individual model. (For a general study of model discriminations at the LHC, see \cite{Hubisz:2008gg}~\cite{Hallenbeck:2008hf}.)

Although we have concentrated on identifying event topologies in this paper, we would like to point out that the functions defined in this paper can also be useful for reducing the combinatorial problems. Once a particular topology is identified. The functions in which this topology has end points can be used to determine the end points of some invariant masses of particles from the same decay chain. Then the end points can be used to cut wrong combinations~\cite{Cheng:2009fw,Blanke:2010cm,Rajaraman:2010hy} by performing an event-by-event analysis and removing combinations with the invariant masses above the corresponding end-point value. The order of the visible particles from a single chain for each individual event remains a difficult problem without additional handles. We also note that the signal events often have peak structure in these functions. The backgrounds, on the other hand, are not expected to have peak structure in general and should be falling rapidly above the kinematic cut. A cut around the peak of the signal region may increase the signal-to-background ratio which could help the discovery and/or the follow-up signal analysis.

In conclusion, we have studied how to identify different event topologies with dark matter particles produced in pairs from cascade decays of heavy particles. Setting $4\,j +\missET$ as a case study, we have shown that one can identify all event topologies based on the existence of end points of several functions of invariant mass distributions defined in this paper. We have also extended our studies to include the cases with ISR and off-shell particles in the decay chains. It is found that most of those topologies can be identified with ${\cal O} (10^3)$ signal events after basic kinematic cuts. We believe that this study should be the {\emph{first}} step towards measuring the masses and spins of the dark matter particles, and similar studies should be performed for events with other final states.

\subsection*{Acknowledgments} 
We would like to thank Jared Kaplan, Joe Lykken, Michael Peskin  and Jay Wacker  for useful discussions and comments. H.-C.~C. thanks the hospitality of Fermilab where this work was initiated. H.-C.~C. is supported by the Department of Energy Grant DE-FG02-91ER40674. SLAC is operated by Stanford University for the US Department of Energy under contract DE-AC02-76SF00515.

%%%%%%%%%%%%%%%%%%%%%%%%%%%%%%%%%%%
\appendix

\section{Invariant Mass End Point Formulae}
\label{sec:endpoint}

\begin{figure}[!ht]
\begin{center}
\includegraphics[width=0.7\textwidth]{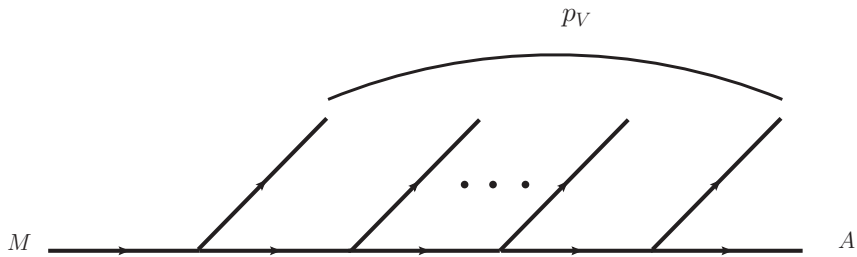} 
\caption{The general Feynman diagram with a mother particle $M$ cascade decays into visible particles and a single missing particle $A$.
}
\label{fig:general_diagram}
\end{center}
\end{figure}

In this Appendix we summarize the formulae for various (up to four particles) invariant mass end points in a decay chain, with either on-shell or off-shell  intermediate particles. We start with a general consideration. Consider that a mother particle $M$ goes through cascade decays by emitting several visible particles and end up with a single missing particle $A$ as shown in Fig.~\ref{fig:general_diagram}.
%%%%
%%%%
Denote the total momentum of all visible particles by $p_V$, the invariant mass-squared of all visible particles is given by
\begin{eqnarray}
p_V^2 = (p_M - p_A)^2 &=& m_M^2 + m_A^2 - 2 p_M \: p_A \nonumber \\
&\to& m_M^2 + m_A^2 - 2 m_M E_A, \quad \mbox{in the rest frame of the mother particle $M$.}
\end{eqnarray}
We see that the maximum of $p_V^2$ occurs when $E_A$ is minimized in the rest frame of $M$. In particular, if it is possible to make $A$ at rest in the $M$ rest frame, then $E_{A, {\rm min}} = m_A$ and $p^2_{V,{\rm max}} = (m_M -m_A)^2$. However, it is not always possible to make $A$ at rest if one of the on-shell decay produces a large boost in some direction which can not be compensated by the boosts in the opposite direction from other stages of the decays. In this case, to reach the maximum of $p^2_{V}$ we still would like to have the other decays to boost $A$ in the opposite direction from the largest-boost decay to minimize $E_A$ in the $M$ rest frame. We will see some examples in the invariant mass end point formulae.

\begin{figure}[!ht]
\begin{center}
\includegraphics[width=0.8\textwidth]{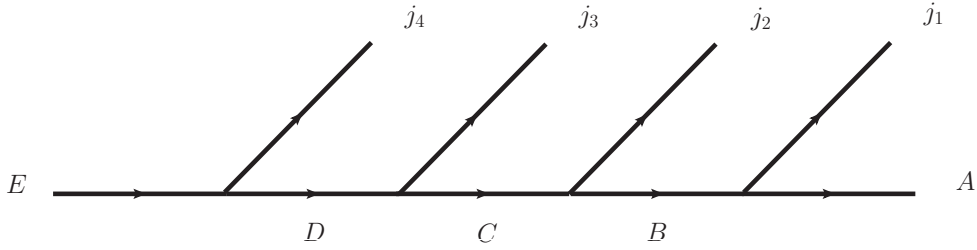} 
\caption{The Feynman diagram with a mother particle $E$ cascade decays into four jets $j_i$ and a single missing particle $A$.
}
\label{fig:4_step_decay}
\end{center}
\end{figure}

Many invariant mass end point formulae  (up to three particles) can be found in the literature~\cite{Hinchliffe:1996iu,Allanach:2000kt,Gjelsten:2004ki,Gjelsten:2005aw,Miller:2005zp,Burns:2009zi,Matchev:2009iw}. For our study, we extend the formulae to any invariant mass combinations in a decay chain with four visible particles, shown in Fig.~\ref{fig:4_step_decay}.
%%%%
%%%%
For simplicity we assume that all visible particles are massless, $p_1^2=p_2^2=p_3^2=p_4^2=0$, which is a good approximation for most cases. The intermediate particles $D,\, C,\, B$ may be off shell. The modifications of the end point formulae for off-shell decays will be remarked when it is relevant. As a rule of thumb, when an end point formula contains a mass parameter explicitly, it applies to the case when the corresponding intermediate particle is on shell. If a mass parameter does not appear in an end point formula, then the formula applies for either that particle being on shell or off shell. We follow the notations in Ref.~\cite{Matchev:2009iw} by defining $R_{ij} \equiv m_i^2/ m_j^2$, where $i,j = A,B,C,D,E$.

\subsection{Two-particle invariant masses}
\begin{enumerate}
\item $m^2_{12,{\rm max}}$:\\
%%%%
\begin{figure}[!ht]
\begin{center}
\includegraphics[width=0.5\textwidth]{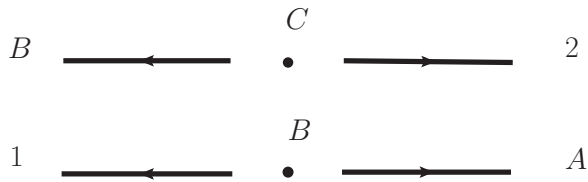} 
\caption{The extremal configuration for $m^2_{12,{\rm max}}$.
}
\label{fig:m12}
\end{center}
\end{figure}
%%%%
The extremal configuration is shown in Fig.~\ref{fig:m12}.
The invariant mass end point for on-shell $B$ is given by the well-known formula
\begin{equation}
m^2_{12,{\rm max}}= \frac{(m_C^2-m_B^2)(m_B^2-m_A^2)}{m_B^2} =m_C^2 (1-R_{BC})(1-R_{AB}).
\end{equation}
In the case that $B$ is off shell, it is possible to make $A$ at rest in the $C$ rest frame, so 
\begin{equation}
m^2_{12,{\rm max}} = (m_C- m_A)^2 = m_C^2 ( 1- \sqrt{R_{AC}})^2, \quad \mbox{if $B$ is off shell.}
\end{equation}
\item $m^2_{13,{\rm max}}$:\\
%%%%
\begin{figure}[!ht]
\begin{center}
\includegraphics[width=0.4\textwidth]{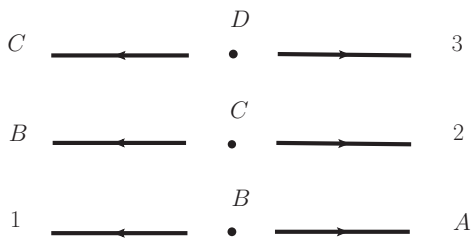} 
\caption{The extremal configuration for $m^2_{13,{\rm max}}$.
}
\label{fig:m13}
\end{center}
\end{figure}
%%%%
The extremal configuration for on-shell decays is shown in Fig.~\ref{fig:m13}.
The corresponding invariant mass end point is
\begin{equation}
m^2_{13,{\rm max}}= \frac{(m_D^2-m_C^2)(m_B^2-m_A^2)}{m_B^2} =m_D^2 (1-R_{CD})(1-R_{AB}).
\end{equation}
If $C$ or $B$ is off shell, then the invariant mass end point occurs in the soft limit of visible particle 2, $p_2 \to 0$. The invariant mass end point formula can be obtained in the same way as $m^2_{12}$,
\begin{equation}
m^2_{13,{\rm max}} = \left\{ \begin{array}{ll}
                                           m_D^2 (1-R_{BD})(1-R_{AB}), & \mbox{if $C$ is off shell,} \\
                                           m_D^2 (1-R_{CD})(1-R_{AC}), & \mbox{if $B$ is off shell.}
                                           \end{array}
                                           \right.
\end{equation}
However, unless the initial particle $D$ is highly boosted, the particle 2 is likely to be too soft to pass the cut in this limit. The end point distribution will not be very sharp on an event sample which includes a hard-enough particle 2. The formula only works as an upper bound in practice and the actual end point may be much smaller depending on the jet $E_T$ cut. Similar consideration also applies to other cases where only one of the two visible particles coming from an off-shell decay is included in the invariant mass calculation. The invariant mass maximum occurs in the soft limit of the other visible particle from that decay. The invariant mass end point is the same as the case with that soft visible particle removed, though in practice it will be smaller and not be as sharp.

\item $m^2_{14,{\rm max}}$:\\
%%%%
\begin{figure}[!ht]
\begin{center}
\includegraphics[width=0.4\textwidth]{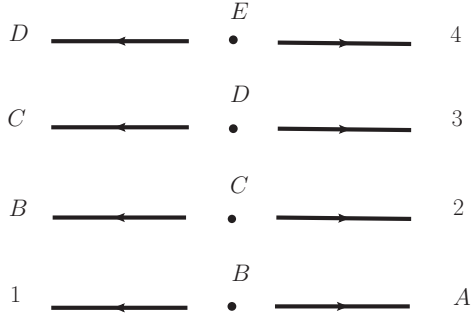} 
\caption{The extremal configuration for $m^2_{14,{\rm max}}$.
}
\label{fig:m14}
\end{center}
\end{figure}
%%%%
The extremal configuration for on-shell decays is shown in Fig.~\ref{fig:m14}.
The corresponding invariant mass end point is
\begin{equation}
m^2_{14,{\rm max}}= m_E^2 (1-R_{DE})(1-R_{AB}).
\end{equation}
When some of the $D,C,B$ intermediate particles are off shell, $m^2_{14,{\rm max}}$ occurs in the soft limit of particle 2 or 3. The configurations become equivalent to the previous cases and one can use the results in these cases with replacements of the appropriate masses.
\end{enumerate}

\subsection{Three-particle invariant masses}
\begin{enumerate}
\item $m^2_{123,{\rm max}}$:\\
First we consider that all three decays are on shell. If $A$ can be put at rest in the $D$ rest frame by combining three boosts, then the end point formula is simply given by $(m_D-m_A)^2$. On the other hand, if one of the decays gives a large boost which can not be compensated by the other two boosts, the end point occurs when the two smaller boosts are in the opposite direction of the largest boost. If one of the intermediate particle is off shell, the two visible particles coming from that three-body decay can have a boost ranging from 0 to the maximal value. Therefore, $A$ can not be put at rest only if the other on-shell decay gives a  boost larger than the maximum boost from the three-body decay.
\begin{enumerate}
\item If $R_{CD} < R_{AC}$, the boost from $D\to C$ decay can not be compensated by the boost from $C \to A$ (irrespective of whether $B$ is on shell or off shell). The end point of  $m^2_{123}$ is given by
\begin{equation}
m^2_{123,{\rm max}} = m_D^2 (1-R_{CD})(1-R_{AC}).
\end{equation}
\item If $R_{BC}< R_{AB}R_{CD}$, the largest boost comes from $C\to B$ decay and the end point  of  $m^2_{123}$ is given by
\begin{equation}
m^2_{123,{\rm max}} = m_D^2 (1-R_{BC})(1-R_{AB}R_{CD}).
\end{equation}
\item If $R_{AB} < R_{BD}$, irrespective of whether $C$ is on shell or not, the end point of  $m^2_{123}$ is given by
\begin{equation}
m^2_{123,{\rm max}} = m_D^2 (1-R_{AB})(1-R_{BD}).
\end{equation}
\item In other cases when there is no single large boost and $A$ can be put at rest in the $D$ rest frame, the end point of $m^2_{123}$ is given by the standard formula,
\begin{equation}
m^2_{123,{\rm max}} = m_D^2 ( 1- \sqrt{R_{AD}})^2.
\end{equation}
\end{enumerate}

\item $m^2_{124,{\rm max}}$:\\
%%%%
\begin{figure}[!ht]
\begin{center}
\includegraphics[width=0.4\textwidth]{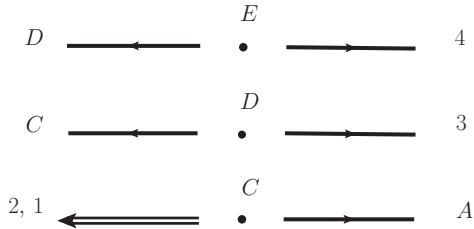} 
\caption{The general extremal configuration for $m^2_{124,{\rm max}}$.
}
\label{fig:m124}
\end{center}
\end{figure}
%%%%
If $C$ or $D$ is off-shell, the end point occurs in the soft limit of visible particle 3. Then it reduces to the previous case (case 1) and the end point formulae can be easily obtained by replacing the appropriate masses. In the following discussion we assume that both $C$ and $D$ are on shell. 
The general extremal configuration is shown in Fig.~\ref{fig:m124}. The double arrow represents the total momentum of particles 1 and 2, while individual particles may go in different directions depending on the relative mass parameters.
\begin{enumerate}
\item If $R_{AC}(1-R_{DE}+R_{CE}) > R_{CE}$
%$R_{AC}(1-R_{DE})^2/(R_{CE}(1-R_{DE}+R_{CE})) > (1-R_{AC})^2$
, the extremal configuration is that particles 1 and 2 move in the same direction, which is opposite to particles 3 and 4.  The end point of $m^2_{124}$ in this case is given by
\begin{equation}
m^2_{124,{\rm max}} = m_E^2 (1-R_{DE})(1-R_{AC}).
\end{equation}
The intermediate particle $B$ can be either on shell or off shell, while other intermediate particles $C$ and $D$ need to be on shell.

\item If all intermediate particles ($B,C,D$) are on shell and $R^2_{BC}(1-R_{DE}+R_{CE}) > R_{AE}$
%$R_{AC}(1-R_{DE})^2/(R_{CE}(1-R_{DE}+R_{CE})) < (R_{AB}-R_{BC})^2$
, and if $R_{AB}>R_{BC}$, the extremal configuration is that particles 1, 3 and 4 move in the same direction which is opposite to particle 2. The end point of $m^2_{124}$ in this case is given by
\begin{equation}
m^2_{124,{\rm max}} = m_E^2 (1-R_{BC}) [1-R_{DE}+R_{CE}(1-R_{AB})].
\end{equation}

\item If all intermediate particles are on shell and $R_{AB}(1-R_{DE}+R_{CE}) > R_{BE}$
%$R_{AC}(1-R_{DE})^2/(R_{CE}(1-R_{DE}+R_{CE})) < (R_{AB}-R_{BC})^2$
, and if $R_{AB}<R_{BC}$, the extremal configuration is that particles 2, 3 and 4 move in the same direction which is opposite to particle 1. The end point of $m^2_{124}$ in this case is given by
\begin{equation}
m^2_{124,{\rm max}} = m_E^2 (1-R_{AB})[1-R_{DE}+R_{CE}(1-R_{BC})].
\end{equation}

\item In other cases, particles 1, 2 and 4 are not collinear in the extremal configuration. The end point is given by
\begin{equation}
m^2_{124,{\rm max}} = m_E^2 (\sqrt{1-R_{DE}+R_{CE}} - \sqrt{R_{AE}})^2.
\end{equation}
\end{enumerate}

\item $m^2_{134,{\rm max}}$:\\
%%%%
\begin{figure}[!ht]
\begin{center}
\includegraphics[width=0.4\textwidth]{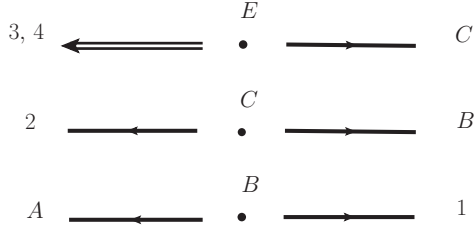} 
\caption{The general extremal configuration for $m^2_{134,{\rm max}}$.
}
\label{fig:m134}
\end{center}
\end{figure}
%%%%
If $B$ or $C$ is off-shell, the end point occurs in the soft limit of visible particle 2. It also reduces to the case 1 and the end point formulae can be easily obtained by replacing the appropriate masses. In the following discussion we assume that both $B$ and $C$ are on shell. 
The general extremal configuration is shown in Fig.~\ref{fig:m134}. The double arrow represents the total momentum of particles 3 and 4, while individual particles may go in different directions depending on the relative mass parameters.
\begin{enumerate}

\item If $\sqrt{R_{BE}}(1-R_{AB}) > (1- R_{CE})$, the extremal configuration is that particles 2, 3 and 4 move in the same direction, which is opposite to particle 1. This is independent whether $D$ is on shell or not. The end point of $m^2_{134}$ in this case is given by
\begin{equation}
m^2_{134,{\rm max}} = m_E^2 (1-R_{CE})(1-R_{AB}).
\end{equation}

\item If all intermediate particles are on shell and $\sqrt{R_{BE}}(1-R_{AB}) < |R_{CD}-R_{DE}|$, and if $R_{CD}>R_{DE}$, the extremal configuration is that particles 1 and 3 move in the same direction, which is opposite to particles 2 and 4. The end point of $m^2_{134}$ in this case is given by
\begin{equation}
m^2_{134,{\rm max}} = m_E^2 (1-R_{DE})(1-R_{CD}R_{AB}).
\end{equation}

\item If all intermediate particles are on shell and $\sqrt{R_{BE}}(1-R_{AB}) < |R_{CD}-R_{DE}|$, and if $R_{CD}<R_{DE}$, the extremal configuration is that particles 1 and 4 move in the same direction, which is opposite to particles 2 and 3. The end point of $m^2_{134}$ in this case is given by
\begin{equation}
m^2_{134,{\rm max}} = m_E^2 (1-R_{CD})(1-R_{DE}R_{AB}).
\end{equation}
\item In other cases, particles 1, 3 and 4 are not collinear in the extremal configuration. The end point is given by
\begin{equation}
m^2_{134,{\rm max}} = m_E^2 (1 - \sqrt{R_{CE} R_{AB}})^2.
\end{equation}

\end{enumerate}

\end{enumerate}
\subsection{Four-particle invariant masses}
\begin{enumerate}
\item $m^2_{1234,{\rm max}}$:\\
Again, the end point depends on whether there is a large boost from one of the on-shell decays which can not be compensated by the combined boost from other decays. 
\begin{enumerate}
\item If $R_{DE} <R_{AD}$ ($B$, $C$ can be on shell or off shell), the end point is given by
\begin{equation}
m^2_{1234,{\rm max}} = m_E^2 (1-R_{DE})(1-R_{AD}).
\end{equation}
\item If $R_{CD} < R_{AC}R_{DE}$ ($B$ can be on shell or off shell), the end point is given by
\begin{equation}
m^2_{1234,{\rm max}} = m_E^2 (1-R_{CD})(1-R_{AC}R_{DE}).
\end{equation}
\item If $R_{BC} < R_{AB}R_{CE}$ ($D$ can be on shell or off shell), the end point is given by
\begin{equation}
m^2_{1234,{\rm max}} = m_E^2 (1-R_{BC})(1-R_{AB}R_{CE}).
\end{equation}
\item If $R_{AB} < R_{BE}$ ($C$, $D$ can be on shell or off shell), the end point is given by
\begin{equation}
m^2_{1234,{\rm max}} = m_E^2 (1-R_{AB})(1-R_{BE}).
\end{equation}
\item In other cases when there is no single large boost and $A$ can be made at rest in the $E$ rest frame, the end point of $m^2_{1234}$ is given by the standard formula,
\begin{equation}
m^2_{1234,{\rm max}} = m_E^2 ( 1- \sqrt{R_{AE}})^2.
\end{equation}
\end{enumerate}

\end{enumerate}

\subsection{End point formulae for functions $F_1 , F_2, F_3, F_4$}

Now we can write down the formulae for the end points (if they exist) for the functions $F_1,F_2,F_3,F_4$ defined in Section~\ref{sec:fourjet} for various topologies.

For the $\fourzero$ topology, the invariant mass end points come from the decay chain with 4 jets. It has exactly the topology in Fig.~\ref{fig:4_step_decay}. The end points for $F_1$--$F_4$ are
\begin{eqnarray}
F_{1,{\rm max}} &=& \sqrt{m^2_{1234,{\rm max}}}, \\
F_{2,{\rm max}} &=& {\rm min} \left\{ \sqrt{m^2_{123,{\rm max}}},\,\sqrt{m^2_{124,{\rm max}}},\,\sqrt{m^2_{134,{\rm max}}},\,\sqrt{m^2_{234,{\rm max}}}\right\} ,\\
F_{3,{\rm max}} &=& \sqrt{m^2_{ij,{\rm max}}}, \, \mbox{ with } {\rm max}\left\{\bigcup_{m,n} m^2_{mn,{\rm max}}\right\}= m^2_{kl,{\rm max}} \mbox{ and } \epsilon^{ijkl} \neq 0, \\
F_{4,{\rm max}} &=& \mbox{min}\left\{   \bigcup_{i, j} \mbox{max} {\Big(} \sqrt{m^2_{ij,\mbox{max}}},\sqrt{m^2_{kl,\mbox{max}}} {\Big)} \right\} \quad \mbox{for} \quad \epsilon^{klij}\neq 0 .
% \mbox{ min } \left\{\bigcup_{m,n} m^2_{mn,{\rm max}}\right\}
\end{eqnarray}

For the $\threeone$ and $\threezeroISR$ topologies, the invariant mass end points come from the decay chain with 3 jets, which are labeled as 1, 2, 3. The end points for $F_1$--$F_4$ are
\begin{eqnarray}
F_{1,{\rm max}} &:&  \mbox{no end point,}\\
F_{2,{\rm max}} &=& \sqrt{m^2_{123,{\rm max}}}, \\
F_{3,{\rm max}} &\leq& {\rm max} \left\{\sqrt{m^2_{12,{\rm max}}},\, \sqrt{m^2_{13,{\rm max}}},\, \sqrt{m^2_{23,{\rm max}}}\right\}, \\
F_{4,{\rm max}} &:& \mbox{no end point.}
\end{eqnarray}
For the three visible particles from the same decay chain, the maximum of $\sqrt{m^2_{12,{\rm max}}}$, $\sqrt{m^2_{13,{\rm max}}}$, $\sqrt{m^2_{23,{\rm max}}}$ occurs when two of the three particles are parallel and the other one is anti-parallel. However, the definition of $F_3$ will take the more energetic one away from the 2 parallel particles to pair with the particle from the other decay chain, so the actual $F_{3,{\rm max}}$ will in general be smaller than the above formula.

For the $\twotwo$ topology, the invariant mass end points can come from either decay chain. We assume that the two decay chains are symmetric and label the two jets on the same chain as 1 and 2. The end points for $F_1$--$F_4$ are
\begin{eqnarray}
F_{1,{\rm max}} &:&  \mbox{no end point,}\\
F_{2,{\rm max}} &:&   \mbox{no end point,}\\
F_{3,{\rm max}} &:&  \mbox{no end point,}\\
F_{4,{\rm max}} &=& \sqrt{m^2_{12,{\rm max}}}.
\end{eqnarray}

The $\twooneISR$ topology is not expected to have a sharp end point in any of $F_1$--$F_4$ functions.

For the parameters used in Sec.~\ref{sec:fourjet}, $m_A=200$~GeV, $m_B=400$~GeV, $m_C=600$~GeV, $m_D=800$~GeV, and $m_E=1000$~GeV, the various end points (in GeV) are listed in Table~\ref{tab:endpoint-theory}.\\
\begin{table}[ht!]
\vspace*{4mm}
\renewcommand{\arraystretch}{1.8}
\centerline{
\begin{tabular}{|c||c|c|c|c|}
\hline
Topology & $F_1$ & $F_2$ & $F_3$ & $F_4$    \\ \hline \hline
$\fourzero$ & 800 & 600 & 394  & 397 \\ \hline
$\threeone$ & -- & 600 & $<458$ & --\\ \hline
$\twotwo$ & -- & -- & -- & 387 \\ \hline \hline
\end{tabular}
}
\caption{Theoretical predictions of the end points in terms of the functions $F_i$ for the spectrum chosen in Sec.~\ref{sec:fourjet}. All numbers are in GeV.}
\label{tab:endpoint-theory}
\end{table}
%%%%%%%%%%%%%%%%%%%%%%%%%%%%%%%%%%%%%%%%%%
\section{Identifying the $\twooneISR$ topology} \label{sec:twooneISR}

To identify the $\twooneISR$ topology, we need to find a way to pick out the ISR jet and distinguish it from the other cases without ISR. The ISR jets should have a falling $E_T$ distribution, while other jets from heavy particle on-shell decays should have an $E_T$ distribution with peak structure if the peak is above the cut. In order to keep the peak structure of those energetic jets, the cut on the jet $E_T$ can not be too strong. Therefore, in this section, we impose a softer cut on the basic kinematics: at least 4 jets with $E_T > 50$~GeV and missing transverse energy $\missET > 100$~GeV. 

\begin{figure}[!ht]
\begin{center}
\includegraphics[width=0.5\textwidth]{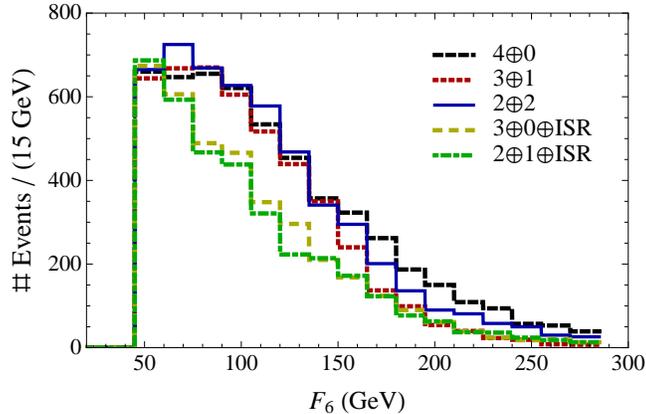} 
\caption{The distributions of $F_6$ for different topologies. We require at least 4 jets with $E_T > 50$~GeV and missing transverse energy $\missET > 100$~GeV. 
}
\label{fig:F6}
\end{center}
\end{figure}

To increase the probability of finding the ISR jet for the $\twooneISR$ topology, we only choose those events in which the two pairs with the largest invariant masses contain the same jet. We then plot the $E_T$ distribution of the remaining one which does not appear in the two largest invariant masses. We use the following function to describe this procedure
\bea
F_6(p_1, p_2, p_3, p_4) &=& E_T(j_l) \,, \nonumber \\
\mbox{such that} && \epsilon^{ijkl}\neq 0 \; \mbox{and} \; m_{ij}, m_{ik}= \mbox{two largest invariant masses.}
\label{eq:F6}
\eea
The event distributions for different topologies in terms of $F_6$ are shown in Fig.~\ref{fig:F6}.
%%%%
%%%%
By choosing a proper bin size, we can see from Fig.~\ref{fig:F6} that the two topologies with ISR have the largest number of events at the first bin, which can be used to distinguish them from topologies without ISR.  

To further distinguish $\twooneISR$ from $\threezeroISR$, we can simply check their patterns of the parameters of the first four functions. 
\begin{table}[ht!]
\vspace*{4mm}
\renewcommand{\arraystretch}{1.7}
\centerline{
\begin{tabular}{|c||cccc||cccc||}
\hline
 & \multicolumn{4}{c||}{bin size = 40 GeV}  & \multicolumn{4}{c||}{bin size = 50 GeV} \\ \hline
   Topologies       & $\frac{b}{ac}(F_1)$ & $\frac{b}{ac}(F_2)$ & $\frac{b}{ac}(F_3)$ & $\frac{b}{ac}(F_4)$ 
                         & $\frac{b}{ac}(F_1)$ & $\frac{b}{ac}(F_2)$ & $\frac{b}{ac}(F_3)$ & $\frac{b}{ac}(F_4)$   
      \\ \hline
$\threezeroISR$ & 0.99 & 0.81 & 1.58 & 0.98 
 & 1.00 & 0.81 & 1.65 & 0.97   
\\ \hline 
$\twooneISR$   &1.11 &1.37  & 1.54 & 1.27
& 1.20 & 1.37  & 2.20 & 1.35    
   \\ \hline \hline
\end{tabular}
}
\label{tab:bacfit30GeVforISR}
\caption{The fitted values of $b/(ac)$, which determine the normalized steepness of the slopes of the histograms after the peak. The broken-line function in Eq.~(\ref{eq:broken-line}) is used to fit distributions. The bin size has been chosen to be 40~GeV (left), and 50~GeV (right). Again each topology has 10000 events before the cuts, and there are 3841 and 3632  events passing the cuts for $\threezeroISR$ and $\twooneISR$, respectively.
}
\end{table}
If $b/(ac) (F_2)$  is smaller than the other values, we can then identify the topology as $\threezeroISR$. On the other hand for the $\twooneISR$ topology, none of the normalized inverse slopes exhibit a particularly small value and $b/(ac) (F_2)$ is not significantly smaller than others. In this case the topology can be $\twooneISR$ or with even more ISRs,

%%%%%%%%%%%%%%%%%%%%%%%%%%%%%%%%%%%
\bibliography{ColliderTopo}
\providecommand{\href}[2]{#2}\begingroup\raggedright\endgroup
%%%%%%%%%%%%%%%%%%%%%%%%%%%%%%%%%%%
\end{document}